\documentclass[twocolumn,prb]{revtex4}

\usepackage{graphicx}
\usepackage{amsmath}
\usepackage{bm}
\usepackage{comment}
\usepackage{multirow}
\usepackage{color}
\usepackage{amsfonts}
\usepackage{ulem}
\newcommand{\red}{}
\newcommand{\blue}{\color{blue}}
\newcommand{\up}{\uparrow}
\newcommand{\down}{\downarrow}
\newcommand{\kv}{{\bf k}}
\newcommand{\DA}[1]{{\color{blue}DA: #1}}

\definecolor{orange}{rgb}{1.0, 0.5, 0.0}
  
\begin{document}
	\title{Altermagnetism from coincident Van Hove singularities: application to $\kappa$-Cl}

\author{Yue Yu$^1$\email{yu36@uwm.edu}, Han Gyeol Suh$^1$\email{jayhansuh@gmail.com}, Mercè Roig$^2$\email{roigserv@uwm.edu}, and Daniel F. Agterberg$^1$\email{agterber@uwm.edu}}
\affiliation{$^1$Department of Physics, University of Wisconsin–Milwaukee, Milwaukee, Wisconsin 53201, USA} 

\affiliation{$^2$Niels Bohr Institute, University of Copenhagen, DK-2100 Copenhagen, Denmark}

\begin{abstract}
Realizing two-dimensional (2D) altermagnets is important for spintronics applications. Here we propose a microscopic template for stabilizing 2D altermagnetism through Van Hove singularities that are coincident in both energy and momentum. These coincident Van Hove singularities are a generic consequence of non-symmorphic symmetries in nine 2D space groups. {\red Due to nontrivial symmetry properties of the Hamiltonian, these coincident Van Hove singularities} allow new hopping interactions between the Van Hove singularities that do not appear in analogous Van Hove singularity based patch models for cuprates and graphene.  We show these new interactions can give rise to various weak coupling, and BCS-based instabilities, including altermagnetism, nematicity, inter-band d-wave superconductivity, and orbital altermagnetic order. We apply our results to quasi-2D organic $\kappa$-Cl in which altermagnetism is known to appear.
\end{abstract}
\maketitle

{\it Introduction: } Altermagnetism, distinct from ferromagnetism and antiferromagnetism, exhibits zero net magnetization with momentum-dependent {\red collinear} spin textures\cite{naka:2019,vsmejkal:2020,vsmejkal:2022emerging,vsmejkal:2022beyond,noda:2016,ahn:2019,hayami:2019,hayami:2020,yuan:2021,mazin:2021,berlijn:2017,zhu:2019,kivelson:2003}. Analogous to unconventional superconductors, it establishes a profound connection between magnetism and topology, hosting nonzero Berry curvature for anomalous Hall transport \cite{vsmejkal:2020,vsmejkal:2022,feng:2022,betancourt:2023,naka:2020,nakatsuji:2015,samanta:2020,surgers:2016,ghimire:2018}. The momentum-dependent spin splitting serves as a intrinsic platform for spin-current coupling\cite{jungwirth:2016}, enabling practical control in spin devices through the application of magnetic fields \cite{vsmejkal:2022giant,hariki:2023}, electrical currents \cite{naka:2019,yuan:2020,gonzalez:2021,naka:2021,shao:2021,bose:2022}, strain \cite{ma:2021,guo:2023,steward:2023,liu:2018,lopez:2012}, torque \cite{karube:2022,bai:2022}, and heat \cite{tomczak:2018,zhou:2023}.
When coupled with superconductors, altermagnetism can induce intriguing phenomena \cite{papaj:2023,beenakker:2023,zhu:2023}, such as the Fulde-Ferrell-Larkin-Ovchinnikov (FFLO) state \cite{zhang:2023,sumita:2023} and new platforms for enabling Majorana particles \cite{ghorashi:2023}. 

As demonstrated through density functional theory (DFT)-based Hartree-Fock calculations \cite{vsmejkal:2022emerging}, the random phase approximation \cite{roig:2024}, and analysis of Fermi surface Pomeranchuk instabilities \cite{Wu:2007,vsmejkal:2022emerging}, altermagnetism is believed to be stabilized through strong on-site Coulomb interactions. Searches for this strong coupling instability form the basis for identifying new altermagnetic materials. In principle, in 2D, Van Hove singularities offer a weak coupling route towards stabilizing altermagnetism. Van Hove singularities are saddle points in the energy dispersion and induce a logarithmic divergence in the 2D density of states. Tuning the chemical potential across such singularities results in divergent susceptibilities, signaling an instability into a variety of possible competing orders that have been examined in cuprates, graphene, and Kagome metals.
\cite{kiesel:2012, nandkishore:2012, wang:2012, furukawa:1998, kampf:2003, honerkamp:2001, le:2009, wang:2013, gonzalez:2008,martin:2008,halboth:2000,honerkamp:2001t,schulz:1987,kiesel:2012s, yu:2012, classen:2020}.  
However, in these applications, altermagnetism is not one of these competing orders, suggesting that this route is not viable for stabilizing this state. {\red Indeed, the closest relative to altermagnetism that has been found is a non-collinear variant that appears only for fine-tuned higher-order Van Hove singularities \cite{classen:2020}.}

{\red In contrast to the strong-coupling mechanisms typically examined with RPA-like approaches}, here we identify a {\red  weak-coupling} mechanism based on usual 2D logarithmic Van-Hove singularities that stabilize altermagnetism. Two ingredients are key to realizing this mechanism. The first is the existence of a pair of Van Hove singularities that are coincident in both energy and momentum. The second, {\red which does not hold for all coincident Van Hove singularites, is  a specific symmetry property of the Hamiltonian, which give rise to} a new hopping interaction between these Van Hove singularities. This hopping interaction does not appear in Van Hove-based patch models for cuprates  \cite{furukawa:1998, kampf:2003, honerkamp:2001, le:2009} and graphene \cite{kiesel:2012, nandkishore:2012, wang:2012, classen:2020} where it is forbidden by translation invariance. Both ingredients are a generic consequence of non-symmorphic symmetries that exist in nine 2D space groups, when spin-orbit coupling is neglected.  Furthermore, unlike the Van Hove driven spin density wave transitions that occur in cuprate \cite{furukawa:1998, kampf:2003, honerkamp:2001, le:2009}, graphene \cite{kiesel:2012, nandkishore:2012, wang:2012, classen:2020}, and Kagome metal \cite{wang:2013} patch models, our altermagnetic mechanism does not require nesting or near-nesting of the bands. Instead, it is based on the BCS instability, which typically yields only a superconducting instability for the other Van Hove singularity scenarios. Our analysis also stabilizes other orders including interband d-wave superconductivity, nematicity, and orbital altermagnetism. To be concrete, we apply our analysis to the organic material $\kappa$-Cl where an altermagnetic state is believed to occur.

{\it Coincident Van Hove singularities:} The phase diagram of the quasi-2D orthorhombic organic compound $\kappa$-(ET)$_2$Cu[N(CN)$_2$]Cl ($\kappa$-Cl, {\red 2D layer group L25 (pba2)}) shares similarities with cuprates \cite{mckenzie:1997}, exhibiting unconventional superconductivity adjacent to magnetism under pressure \cite{williams:1990,miyagawa:2004,kagawa:2005} and anion substitution \cite{kurosaki:2005,miyagawa:2002,kanoda:1997,mayaffre:1995,de:1995,kanoda:1996,urayama:1988,kini:1990}. {\red Previous theoretical studies, largely based on strong coupling RPA-like calculations, reveal superconducting and altermagetic states\cite{kuroki:2002,sekine:2013,guterding:2016}.} DFT calculations reveal a Van Hove singularity at the $S=(\pi,\pi)$ point \cite{koretsune:2014}. Recently, considerable hole-doping has been achieved in $\kappa$-Cl \cite{kawasugi:2016}. {\red As the Van Hove singularity is approached, a substantial reduction in spectral weight is observed, revealing the importance of electronic correlations\cite{kawasugi:2016}.}  Here we show that this Van Hove singularity consists of a pair of Van Hove singularities that are coincident in both energy and momentum, {\red and also satisfies the symmetry conditions that allow the new hopping interaction mentioned above. We also show that in the vicinity of the Van Hove singularity, interactions that fall outside of RPA-like approaches drive a weak-coupling instability to altermagnetism and other novel states. }


\begin{figure}[ht]
\centering
\includegraphics[height=5cm]{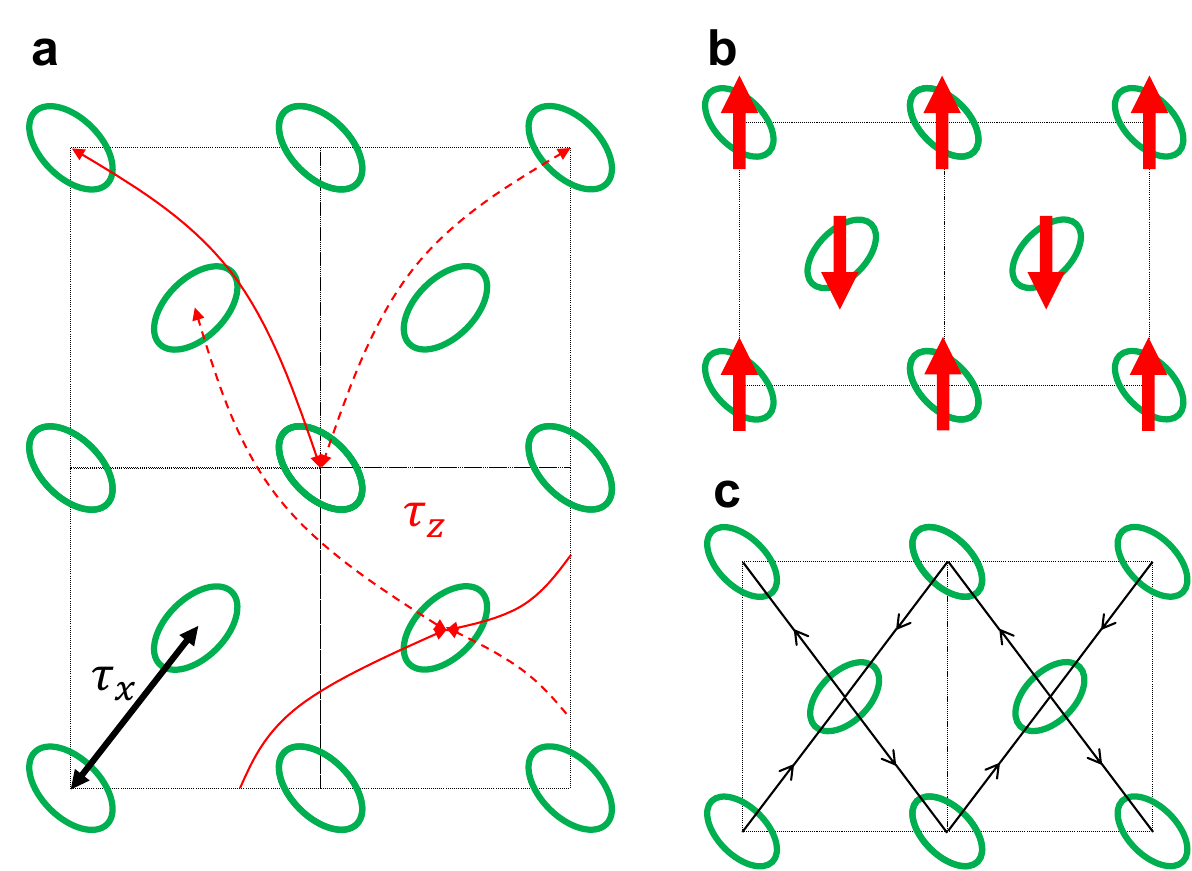}
\caption{Crystal structure. (a) {\red Hopping parameters. (b) Altermagnet where arrows represent spins. (c) Orbital altermagnet where arrows represent currents (even-parity current loop order).  We name this state an orbital altermagnet since the local moments induced by the current loops form an altermagnet (which has  the same symmetries as the altermagnet depicted in the top right).}}
\label{F:crystal}
\end{figure}

{\red For pedagogical purposes, we start with a tight-binding Hamiltonian. A general group theory analysis can be found later.} $\kappa$-Cl has four ET molecules per unit cell, forming two molecule-dimers (Fig.~\ref{F:crystal}) at positions: $(r_1,r_2)=\{(0,0),(1/2,1/2)\}$. As a minimal model capturing the coincident Van Hove singularities, we consider one orbital on each dimer position. The Pauli matrices $\tau_i$ act in dimer (or sublattice) space and the Pauli matrices $\sigma_i$ act in spin space. While spin-orbit coupling (SOC) is negligible \cite{koretsune:2014,naka:2019}, the normal state tight-binding Hamiltonian is:
\begin{equation}
\begin{split}
H&=2t_1\cos{k_x}+2t_2\cos k_y+4t_3\cos k_x \cos k_y\\
&+4t_4 \cos\frac{k_x}{2}\cos\frac{k_y}{2}\tau_x+t_5 \sin k_x\sin k_y \tau_z-\mu.
\label{eq:TBH_RuO2}
\end{split}
\end{equation}
{\red Here, $\tau_x$ term describes inter-sublattice hopping (thick black arrows in Fig.~\ref{F:crystal}). $\tau_z$ term captures the difference in intra-sublattice hopping between $(1,1)$ and $(1,-1)$ directions (solid/dashed red arrows in Fig.~\ref{F:crystal}), which is opposite on the two sublattices.} Each band is doubly degenerate, consisting of spin-up and down states. Due to the nonsymmorphic symmetries, the band without SOC is 4-fold degenerate at the entire Brillouin zone boundary $k_x=\pi$ and $k_y=\pi$ \cite{suh:2023}. Near the Brillouin zone corner S=$(\pi,\pi)$, {\red where the coincident Van Hove singularities appear}, we obtain the kp Hamiltonian:
\begin{equation}
H(\pi+k_x,\pi+k_y)=t_xk_x^2+t_yk_y^2 +k_xk_y(t_4\tau_x+t_5\tau_z)-\widetilde{\mu},
\label{Eq:kp}
\end{equation}
with $\widetilde{\mu}=\mu+2t_1+2t_2-4t_3$, $t_x=(t_1-2t_3)$, and $t_y=(t_2-2t_3)$. {\red This expansion reveals a central property of all the theories we examine here, both the operators $\tau_x$ and $\tau_z$ are multiplied by the momentum function $k_xk_y$. Since the Hamiltonian must be invariant under all orthorhombic symmetries, this implies that {  both} $\tau_x$ and $\tau_z$ share the same symmetry as $k_xk_y$ at the $S$ point. A group theory analysis (given in detail later) applied to the momentum points and 2D and 3D space groups in Table 1 reveals that the kp-theory of Eq.~\ref{Eq:kp} also appears in these cases and has the most general form allowed by symmetry (for the 3D groups, a term $a_zk_z^2$ is also allowed, and the physics discussed here occurs in the quasi-2D limit, where  $a_z$ is small relative to $t_x$ and $t_y$). 
The other non-trivial sublattice operator $\tau_y$ is time-reversal odd, {\red but invariant under all crystal symmetries. $\tau_y$ can only appear in a Hamiltonian multiplied by spin-1/2 operators \cite{suh:2023} as a SOC term.} Here we ignore SOC and include a discussion of its effects in the appendix.}

The resulting dispersions of the two bands are:
\begin{equation}
\begin{split}
E_{1,2}=t_xk_x^2+t_yk_y^2\pm\sqrt{t_4^2+t_5^2}k_xk_y-\widetilde{\mu}.
\end{split}
\label{Eq:dispersion}
\end{equation}
The saddle point conditions for these two dispersions are the same: $t_4^2+t_5^2+(t_x-t_y)^2>(t_x+t_y)^2$. Consequently, each band hosts a Van Hove singularity at the band crossing point $S=(\pi,\pi)$.



In the following, we work in the band basis for which the kp Hamiltonian, Eq.~\ref{Eq:kp} is diagonal. To diagonalize the kp Hamiltonian, we take a {\red k-independent} unitary transformation $u=\cos(\theta/2) I-i\sin(\theta/2)\tau_y$ where $\cos\theta=t_5/\sqrt{t_4^2+t_5^2}$. {\red In the band basis, $\widetilde{\tau_y}$ is unchanged from $\tau_y$ in the sublattice basis. $\widetilde{\tau_x}$ and $\widetilde{\tau_z}$ in the band basis are linear combinations of 
$\tau_x$ and $\tau_z$, with  $\widetilde{\tau_x}=\cos\theta\tau_x-\sin\theta\tau_z$ and $\widetilde{\tau_z}=\sin\theta\tau_x+\cos\theta\tau_z$. Since $\tau_{x,z}$ have the same symmetry and $\theta$ is a constant, $\widetilde{\tau_{x,z}}$ should remain the same symmetry as $k_xk_y$. As interactions can be constructed from multiplying Fermionic bilinears with the same symmetry, this indicates the existence of a new interaction term with respect to kp theories in which $\tau_x$ and $\tau_z$ have different symmetries.}

\begin{figure}[h]
\centering
\includegraphics[width=6cm]{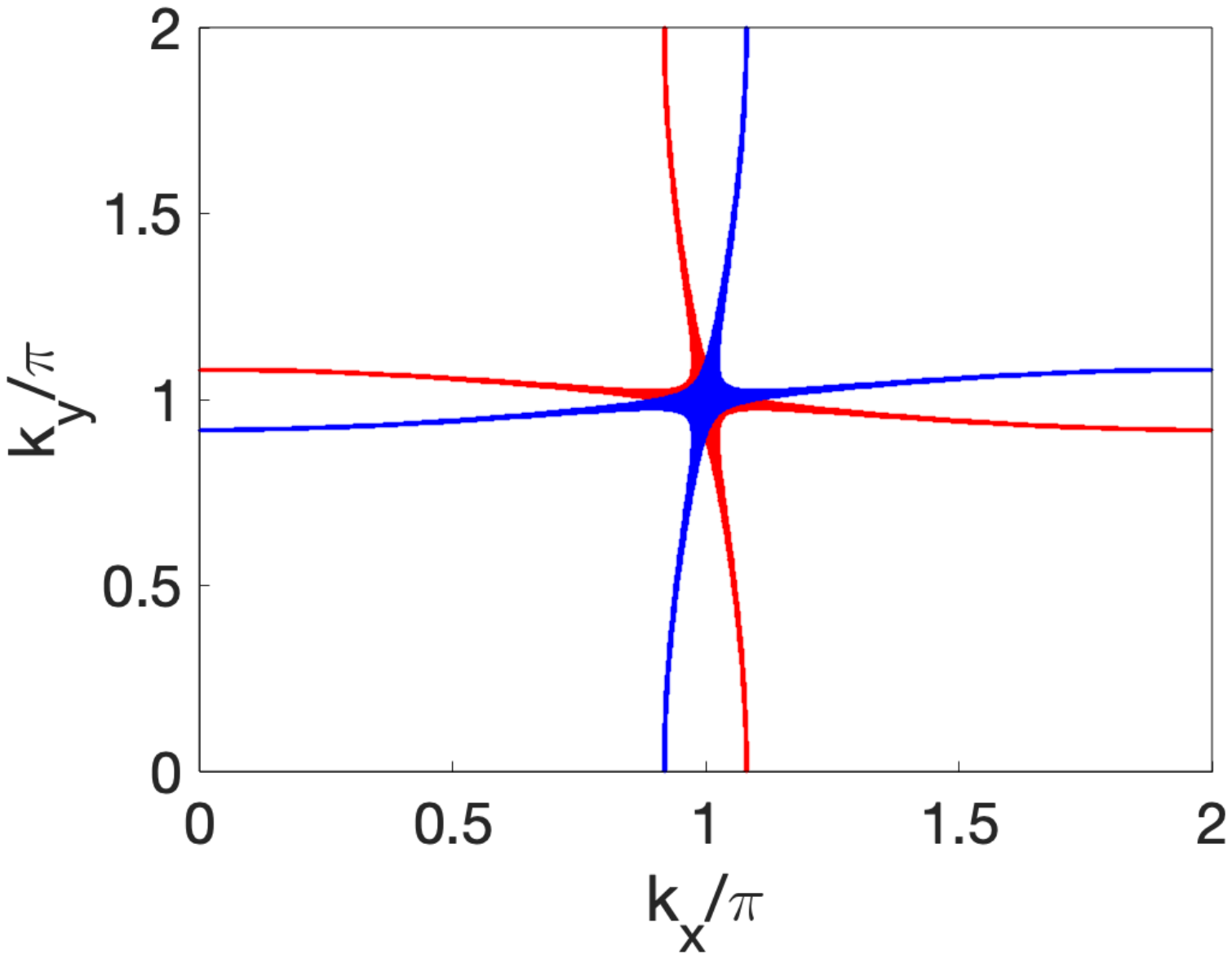}
\caption{{\red Density of states is concentrated near the band crossing point $(\pi,\pi)$, where the two bands are shown in different colors. In the patch model, states on the red band are replaced by a single point at $(\pi,\pi)$, while states on the blue band are replaced by another point at $(\pi,\pi)$. Here, we take $t_1=t_2=t_5=1$, $t_3=0$, $t_4=8$, and $\mu=-4$. $2000^2$ points are sampled, and states with energy ($|E|<0.1$) are plotted.}
}
\label{F:FS}
\end{figure}

\begin{figure}[ht]
\centering
\includegraphics[width=8cm]{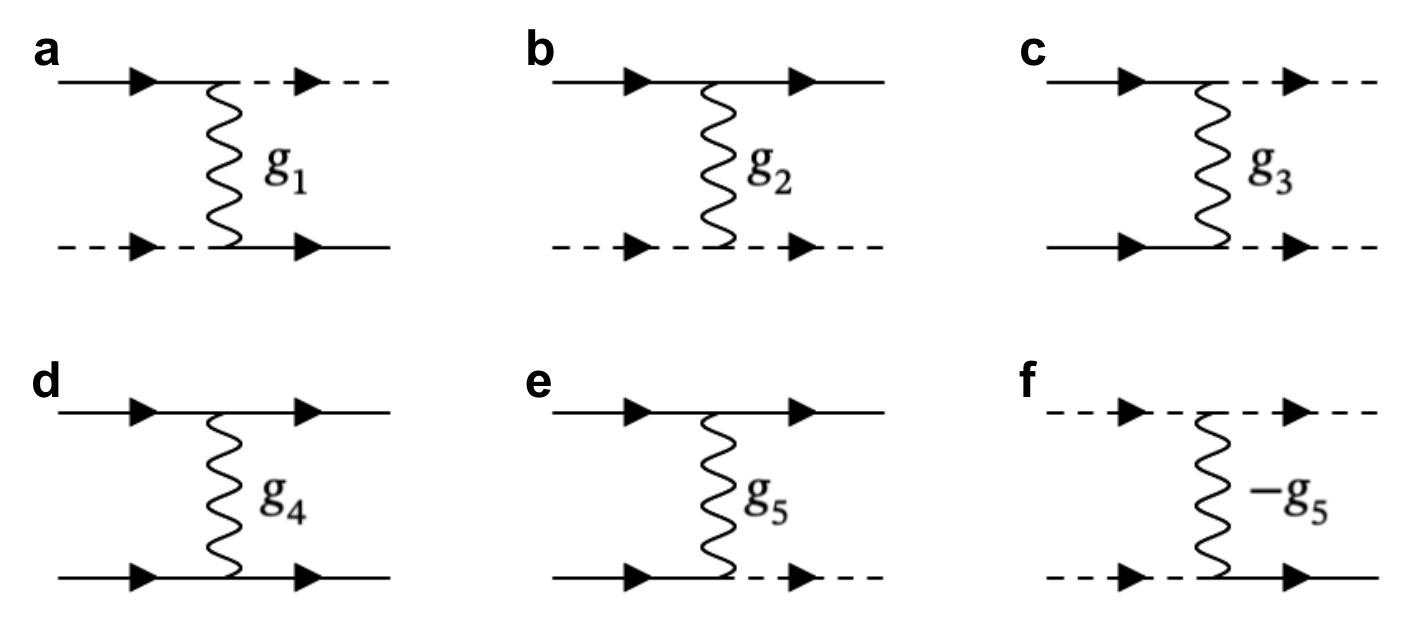}
\caption{Allowed interactions for two coincident  patches. (a-f) the exchange interaction $g_1$, inter-patch Hubbard interaction $g_2$, pair-hopping interaction $g_3$, intra-patch Hubbard interaction $g_4$, and the new hopping interaction $g_5$. Here, upper propagators are for spin-up, while lower ones are for spin-down. {\red Solid\&dashed lines label band 1\&2.}} 
\label{F:interactions}
\end{figure}

{\it Patch model and interactions:} The coincident Van Hove singularities provide a novel platform to study strongly interacting fermions and a natural mechanism for stabilizing 2D altermagnetism. To understand this, it is useful to consider a patch model \cite{furukawa:1998}. Here, the density of states (DOS) of each band is approximated as a patch situated solely at the Van Hove point. {\red With a Van Hove singularity on each band, we end up with two {\red DOS} patches located at the same point: $(\pi,\pi)$, as shown in Fig.\ref{F:FS}}. This patch model has similarities with that examined for cuprates. For cuprates, there are also two Van Hove singularities, but these Van Hove singularities are not coincident and appear at the two distinct momenta $(0,\pi)$ and $(\pi,0)$. The key difference between patch models for coincident Van Hove points and those for the cuprates is in the interactions allowed by symmetry. Both models contain the standard interactions $g_1$, $g_2$, $g_3$, and $g_4$. However, our coincident Van Hove patch model contains a new hopping interaction, $g_5$, shown in Fig.\ref{F:interactions}

This $g_5$-interaction involves a hopping between the two {\red bands (DOS patches)} which, due to momentum conservation, is permissible only when two patches coincide. 
{\red The existence of coincident Van Hove singularities is not sufficient for the existence of the $g_5$ interaction. Additionally, 
the Hamiltonian must satisfy the symmetry condition discussed just below Eq.~2 that  $\widetilde{\tau}_x$ and $\widetilde{\tau}_z$ have the same symmetry. To demonstrate this, we use Cooper pair operators to derive the $g_5$ interaction. Cooper pair operators are convenient since they automatically encode the Pauli exclusion principle. The interactions in Fig.\ref{F:interactions} involve annihilating two electrons (a Cooper pair) from the left and then creating another Cooper pair with the same symmetry to the right. For $g_5$, the corresponding Cooper pairs are $\widetilde{\tau_x}i\sigma_y:\Delta_{xy,1}=c_{2\downarrow}c_{1\uparrow}+c_{1\downarrow}c_{2\uparrow}$ and $\widetilde{\tau_z}i\sigma_y: \Delta_{xy,2}=c_{1\downarrow}c_{1\uparrow}-c_{2\downarrow}c_{2\uparrow}$  (here 1 and 2 in Fermionic operators label the bands). Since these two Cooper pairs share the same symmetry,  the interaction $g_5\Delta_{xy,1}^\dagger\Delta_{xy,2}$ is symmetry allowed. If the $\tilde{\tau}_x$ and $\tilde{\tau}_z$ operators had different symmetries, this interaction would be symmetry forbidden.} Expanding $g_5\Delta_{xy,1}^\dagger\Delta_{xy,2}$ yields
\begin{equation}
\begin{split}
&H^{int}_{g_5}=g_5(c_{2\downarrow}c_{1\uparrow}+c_{1\downarrow}c_{2\uparrow})^\dagger (c_{1\downarrow}c_{1\uparrow}-c_{2\downarrow}c_{2\uparrow})+h.c.
\end{split}
\end{equation}
{\red In the two $g_5$ panels in Fig.~3, the two input electrons from the left annihilates the Cooper pair $\Delta_{xy,2}$, and the two output electrons to the right creates the Cooper pair $\Delta^\dagger_{xy,1}$. The sign difference on the two bands in $\Delta_{xy,2}$ results in the sign difference between the two $\pm g_5$ panels.}

The $g_5$ interaction can be generated by an on-site Hubbard interaction after transforming to the band basis, so its magnitude can be large. In particular, for the interaction 
\begin{equation}
    H_{\rm int} = U \sum_{m} n_{m \uparrow} n_{m \downarrow},
\end{equation}
where $m$ denotes the sublattice index, 
we find {\red its components in the band basis} (denoted by $g_i^0$):
$g^0_1=g^0_2=g^0_3=(U/2)\sin^2\theta$, $g_3^0+g^0_4=U$, $g^0_4-g^0_3=U\cos^2\theta$, and $g^0_5=(U/2)\sin\theta\cos\theta$. {\red Recall $\cos\theta\equiv t_5/\sqrt{t_4^2+t_5^2}$, the existence of $g_5$ thus requires both $k_xk_y(t_4\tau_x+t_5\tau_z)$ terms in the dispersion. In single-band cuprates, even if we fold the Brillouin zone such that the two Van Hove singularities coincide, $g_5$ will still vanish as it corresponds to $\theta=\pi/2$. }


When the chemical potential is tuned close to the Van Hove singularity, the enhancement in the density of states results in substantial corrections to interaction strengths. These corrections originate through the intra- and inter-band particle-particle ($\chi^{intra}_{pp}$, $\chi^{inter}_{pp}$ ) and particle-hole ($\chi^{intra}_{ph}$, $\chi^{intra}_{ph}$ ) susceptibilities for the free fermions \cite{kiesel:2012, nandkishore:2012, wang:2012, furukawa:1998, kampf:2003, honerkamp:2001, le:2009, wang:2013, gonzalez:2008,martin:2008,halboth:2000,honerkamp:2001t,schulz:1987,kiesel:2012s, yu:2012}. {\red Here, these susceptibilities are 
\begin{equation}
\begin{split}
&\chi_{ph}^{intra}=-\lim_{{\bf q}\to 0}\frac{1}{N}\sum_{\bf k}\frac{f[E_1({\bf k})]-f[E_1({\bf k+q})]}{E_1({\bf k})-E_1({\bf k+q})}\\
&\chi_{ph}^{inter}=-\lim_{{\bf q}\to 0}\frac{1}{N}\sum_{\bf k}\frac{f[E_1({\bf k})]-f[E_2({\bf k+q})]}{E_1({\bf k})-E_2({\bf k+q})}\\
&\chi_{pp}^{intra}=-\lim_{{\bf q}\to 0}\frac{1}{N}\sum_{\bf k}\frac{f[E_1(-{\bf k}))-f(-E_1({\bf k+q})]}{E_1(-{\bf k})+E_1({\bf k+q})}\\
&\chi_{pp}^{inter}=-\lim_{{\bf q}\to 0}\frac{1}{N}\sum_{\bf k}\frac{f[E_1(-{\bf k})]-f[-E_2({\bf k+q})]}{E_1(-{\bf k})+E_2({\bf k+q})},
\end{split}
\end{equation}
where $f[E]$ is the Fermi Dirac distribution.} 
The dominant corrections stem from $\chi^{intra}_{pp}$.  This is because $\chi^{intra}_{pp}$ exhibits the conventional BCS logarithmic divergence multiplied by the logarithmic divergence in the density of states ($\chi^{intra}_{pp}$ diverges as $\log^2\Lambda/T$) while other susceptibilities only exhibit the latter and hence diverge as $\log\Lambda/T$. Keeping only the dominant interaction corrections, the  one-loop BCS corrections to the interactions are:
\begin{equation}
\begin{split}
&\Delta g_1=-2\chi_{pp}^{intra}g_5^2\\
&\Delta g_2=-2\chi_{pp}^{intra}g_5^2\\
&\Delta g_3=-2\chi_{pp}^{intra}g_3g_4\\
&\Delta g_4=-\chi_{pp}^{intra}(g_3^2+g_4^2)\\
&\Delta g_5=-\chi_{pp}^{intra}(g_4-g_3)g_5\\
\rightarrow
&\left[\begin{array}{l}
     \Delta (g_4+g_3)=-\chi_{pp}^{intra}(g_4+g_3)^2 \\
     \Delta (g_4-g_3)=-\chi_{pp}^{intra}(g_4-g_3)^2 
\end{array}\right].
\end{split}
\label{Eq:6}
\end{equation}
Let us start with $g_5=0$, then these equations are the same as those found in cuprate patch models \cite{furukawa:1998, honerkamp:2001}. Hence, a bare attractive $g^0_4+g^0_3$ (or $g^0_4-g^0_3$) enhances itself and gives an intra-band s-wave superconducting $\Delta=i\sigma_y$ (or d-wave superconducting $\Delta=i\widetilde{\tau_z}\sigma_y$) instability. For onsite Hubbard interactions {\red and generic $\theta$}, we have $g^0_4>g^0_3>0$, causing the BCS correction to suppress these two instabilities.

\begin{figure}[ht]
\centering
\includegraphics[width=7cm]{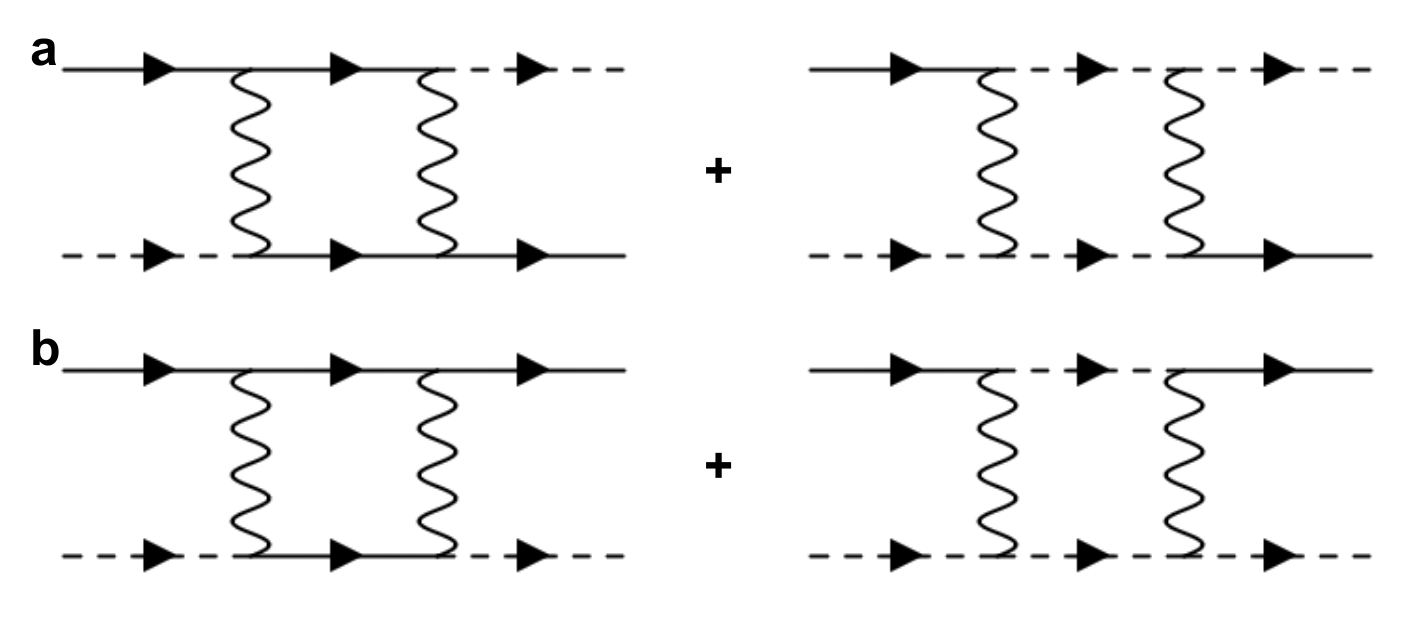}
\caption{$g_5$-interaction leads to intra-band particle-particle corrections to {\red (a)} $g_1$ and {\red (b)} $g_2$.}
\label{F:1loop2}
\end{figure}

The $g_5$-interaction introduces new one-loop BCS corrections to the exchange interaction $g_1$, and the inter-band Hubbard interaction $g_2$, as shown in Fig.\ref{F:1loop2}. {\red The first diagram in panel (a) illustrates a correction to the exchange interaction $g_1$. This correction involves two $g_5$ interactions. This diagram features two internal solid lines moving in the same direction, which are evaluated to be the negative BCS susceptibility, $-\chi^{intra}_{pp}$, from the first band. Similarly, the second diagram has two $-g_5$ interaction lines and a negative BCS susceptibility from the second band. These two diagrams thus add up to $\Delta g_1=-2\chi_{pp}^{intra}g_5^2$ in Eq.\ref{Eq:6}. The two diagrams in panel (b) add up to $\Delta g_2=-2\chi_{pp}^{intra}g_5^2$. Within BCS corrections, $g_5$ does not affect $g_{3,4}$. }

Notably, these corrections are always negative, regardless of the signs of interactions. This can drive $g_{1,2}<0$. In the appendix, we consider a patch renormalization group study including subleading corrections, which are important as $g_4-g_3$ vanish \cite{arovas:2022}. We consider the small $g_{3,4}$ sector and explicitly show that BCS corrections from $g_5$ can push the RG flow to stable fixed points with divergent $g_{1,2}<0$.

\begin{table}[ht]
\begin{tabular}{|c|c|}
\hline
Instability & eigenvalue \\ \hline 
d-wave altermagnetism: $\widetilde{\tau_z}\sigma\sim k_xk_y\sigma$ & $\chi_{ph}^{intra}|g_2|$ \\ 
nematicity: $\widetilde{\tau_x}\sim k_xk_y$ & $\chi_{ph}^{inter}|g_1|$\\  
orbital altermagnetism: $\widetilde{\tau_y}$ & $\chi_{ph}^{inter}|g_1|$\\
d-wave SC: $\widetilde{\tau_x}(i\sigma_y)\sim k_xk_y(i\sigma_y)$ &$\chi_{pp}^{inter}|g_1+g_2|$\\ \hline
\end{tabular}
\caption{Orders stabilized by coincident Van Hove points and corresponding {\red symmetries and} eigenvalues. }
\label{T:1}
\end{table}

{\it Competing instabilities:}
The above discussion reveals that negative $g_1$ and $g_2$ can become the dominant interactions, and so we may neglect other interactions. We find that $g_1$ and $g_2$ generically lead to competing instabilities in the intra-band particle-hole, inter-band particle-hole, and inter-band particle-particle channels (Table.\ref{T:1}). 
{\red These instabilities are obtained from self-consistent one-loop vertex corrections in the appendix, and here are illustrated from a mean-field perspective. Specifically, d-wave altermagnetism $\widetilde{\tau_z}\sigma$ requires an antiferromagnetic interaction between the two bands, which is contributed by the attractive inter-band Hubbard interaction $g_2$ (marked in blue below):
\begin{equation}
\begin{split}
&S_{1z}S_{2z}=(n_{1\uparrow}-n_{1\downarrow})(n_{2\uparrow}-n_{2\downarrow})\\
=&{\blue -(n_{1\uparrow}n_{2\downarrow}+n_{2\uparrow}n_{1\downarrow})}+n_{1\uparrow}n_{2\uparrow}+n_{1\downarrow}n_{2\downarrow}
\end{split}
\end{equation}
The altermagnetism is in the intra-band particle-hole channel, so the instability criteria is $\chi_{ph}^{intra}|g_2|>1$, as shown in Table.\ref{T:1}.
Other} instabilities include inter-band $d_{xy}$ superconductivity, $\epsilon_{xy}$ nematicity, and orbital altermagnetic order. {\red These order parameters inherit non-trivial symmetries through the band degrees of freedom $\widetilde{\tau_i}$, even though the coincident Van Hove singularity is located at a single momentum point. The orbital altermagnetic order $\widetilde{\tau_y}$ is a current-loop order breaking time-reversal symmetry while preserving all crystal reflection symmetries (Fig.\ref{F:crystal}).} This state will induce an anomalous Hall effect under the application of an $\epsilon_{xy}$ strain, even without the presence of SOC (a detailed explanation is in the appendix). When SOC is present, this current loop order and the conventional $d$-wave {\red spin} altermagnetic order {\red $\widetilde{\tau_z}\sigma_z$ will coexist as these two orders share the same symmetry.}  It is notable that the d-wave superconductivity we find is inter-band, and cannot be stabilized through a conventional BCS mechanism.

\begin{figure}[ht]
\centering
\includegraphics[width=6cm]{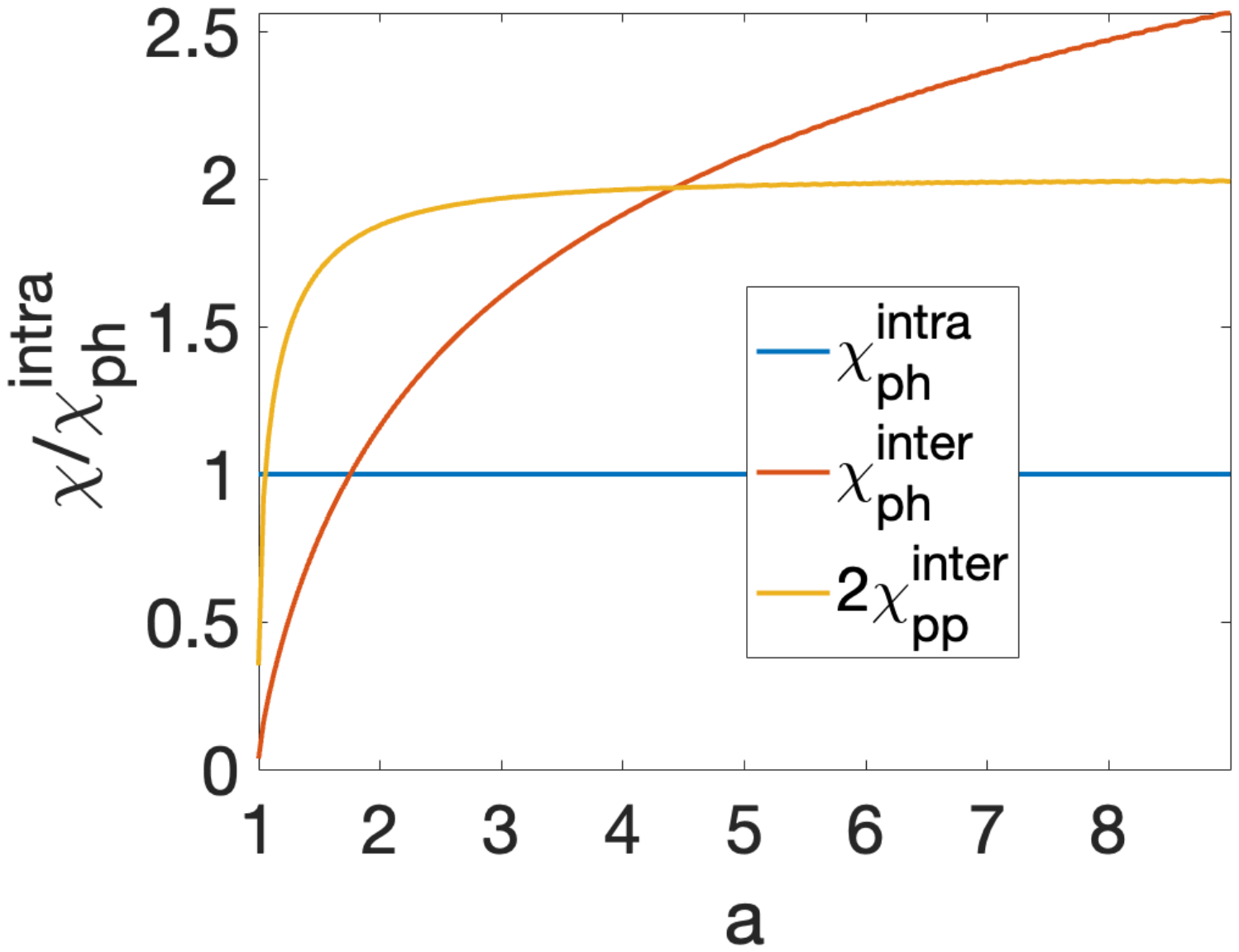}
\caption{Susceptibilities $\chi_{pp}^{inter}$, $\chi_{ph}^{inter}$ and $\chi_{ph}^{intra}$, normalized with respect to $\chi_{ph}^{intra}$, with $t_x=t_y$ and $a\equiv\frac{\sqrt{t_4^2+t_5^2}}{2t_x}$. $2\chi_{pp}^{inter}$ is plotted as the SC vertex is contributed by both $g_1$ and $g_2$. {\red $N=2000^2$ points are sampled, for $k_{x,y}\in[-5,5]$. Temperature $T/t_x=0.01$ is taken. ${\bf q}=(10^{-4},0)$ is used in $\chi_{ph}^{intra}$. ${\bf q}=(0,0)$ is used in $\chi_{ph}^{inter}$ and $\chi_{pp}^{inter}$.}} 
\label{F:susceptibilies}
\end{figure}

While the one-loop correction for the interaction comes from the dominant intra-band particle-particle (BCS) channel, the one-loop correction for the vertices all belong to the three subleading channels.
The ultimate leading instability can be determined by the magnitude of the three eigenvalues $\chi_{pp}^{inter}|g_1+g_2|$, $\chi_{ph}^{inter}|g_1|$ and $\chi_{ph}^{intra}|g_2|$. The three susceptibilities share the same scaling: $\chi\propto\log \Lambda_i/T$, generically with different energy cutoff $\Lambda_i$.
To examine what parameter range enables the different instabilities, we set the cutoffs in all three susceptibilities the same and consider the limit $g_1=g_2$.  This limit occurs when only BCS corrections from on-site Coulomb interactions are included but generally $g_1\ne g_2$. The results then depend upon the kp dispersion in Eq.\ref{Eq:dispersion}. 
Fig.\ref{F:susceptibilies} reveals that all instability channels are stable for some choice of $a\equiv\frac{\sqrt{t_4^2+t_5^2}}{2t_x}$ (Van Hove dispersion requires $a>1$). We note that once $g_1>g_2$ (or $g_1<g_2$), the inter-band (or intra-band) altermagnetic phase will expand.

\begin{figure}[ht]
\centering
\includegraphics[width=8cm]{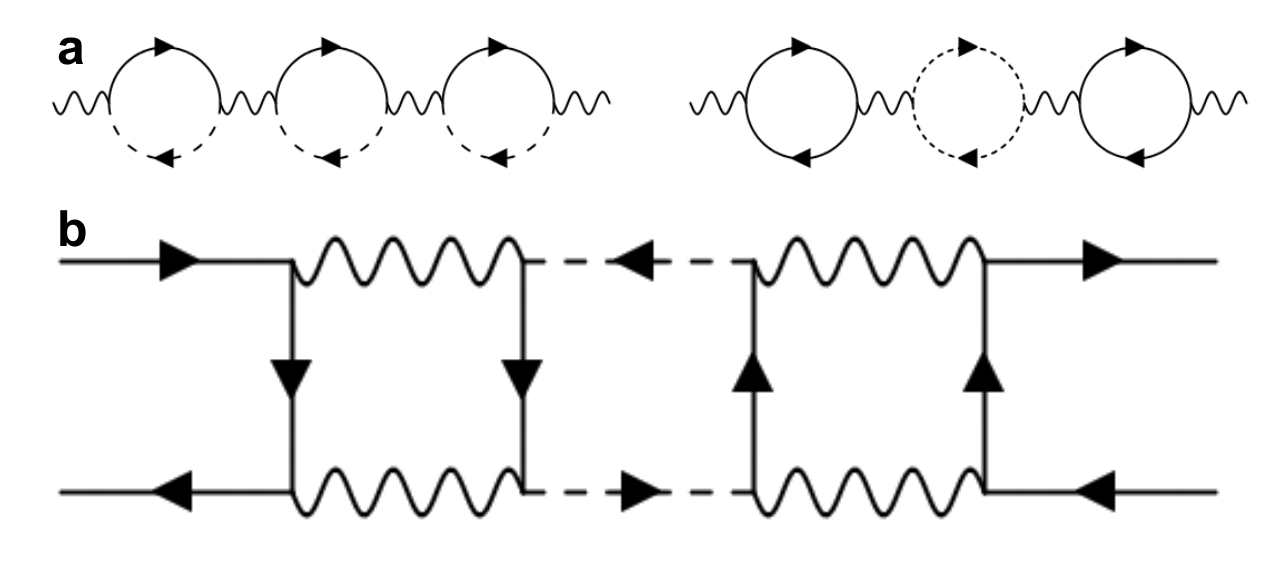}
\caption{Comparison of conventional and the $g_5$ bubbles to magnetism. (a) Conventional Stoner bubbles from $g_{1,2}$, for intra-band and inter-band particle-hole channel. (b) Contribution through intra-band particle-particle channel from the $g_5$ interaction. The two pairs of vertical propagators can be from both patches. }
\label{F:7}
\end{figure}

It is worth noting that the above results are not captured by the standard random phase approximation (RPA). The RPA diagram is shown in the panel (a) in Fig. \ref{F:7}. It gives the usual instability criteria $|g_2|\chi_{ph}^{intra}>1$ (topleft) and $|g_1|\chi_{ph}^{inter}>1$ (topright). The BCS instability for altermagnetism is equivalent to a two-loop correction described by the panel (b). {\red This diagram is derived from panel (a) by substituting the $g_2$ interaction with its $g_5$-correction in Fig.\ref{F:1loop2}. The instability criteria is thus $|\Delta g_2|\chi_{ph}^{intra}=2|g_5|^2\chi_{pp}^{intra}\chi_{ph}^{intra}>1$.}
Since $\chi_{pp}^{intra}\gg\chi_{ph}^{intra},\chi_{ph}^{inter}$ at low-temperature, the BCS correction from $g_5$ describes a stronger weak-coupling instability. Since the $g_5$ contributions exceed the RPA-based contributions in the weak-coupling limit, it is reasonable to include them beyond the conventional RPA-based contributions when addressing intermediate and strong coupling problems in future studies.

{\red Our coincident Van Hove mechanism relies on the existence of a $g_5$ hopping interaction and  exclusively yields $d$-wave altermagnetic states. Given the discovery of $g$-wave altermagnetism \cite{krempasky:2024,mazin:2023}, it is interesting to ask if there exists a weak-coupling Van Hove scenario that stabilizes such a $g$-wave state. In the appendix we show that this is indeed possible for a different coincident Van Hove scenario that also applies to many of the space groups examined here. In this case, a momentum dependent spin-splitting of the form $k_xk_y(k_x^2-k_y^2)$ is generated from coincident Van Hove singularities with a symmetry that forbid the existence of the $g_5$ interaction. This Van Hove scenario can be mapped onto the patch model for the cuprates\cite{schulz:1987} and this mapping reveals that the Neel spin density wave state found for the cuprate case maps to the $g$-wave altermagnetic state. In contrast to the $d$-wave altermagnetic that appears when $g_5\ne 0$, the $g$-wave altermagnetic state requires nesting to become stable.}

{\red 
{\it Group theory arguments:} 
We now explicitly derive the symmetry requirements that underlie the kp Hamiltonian in Eq.\ref{Eq:kp}. Key to this argument are non-symmorphic 2-fold symmetries. For these symmetries, we adopt the notation  $\widetilde{O}=\{O|\bf{t}\}$ to describe a reflection symmetry $O$ followed by a half-translation vector $\bf{t}$. We also denote a pure translation as $\{t^x,t^y\}$. The first requirement for Eq.\ref{Eq:kp} is a 2-fold sublattice degeneracy, which appears independently of 2-fold spin degeneracy. This degeneracy is ensured by the product of spinless time-reversal symmetry $T$ (here $T$ is represented by the complex conjugation operator) and a non-symmorphic symmetry. For example in $\kappa$-Cl, reflection symmetry $\widetilde{M_x}=\{M_x|1/2,1/2\}$ takes the position $(x,y)$ to $(-x+1/2,y+1/2)$. We have $(T\widetilde{M_x})^2=-1$, which follows for $T^2=1$ and $(\widetilde{M_x})^2=\{0,1\}=\exp(ik_y)$ when the momentum of the Bloch state has $k_y=\pi$. The TRIM point thus exhibits 2-fold sublattice degeneracy. In this work, we do not consider TRIM points with higher sublattice degeneracies.

As the $T$ invariant operators $\tau_x$ and $\tau_z$ both share the same non-trivial transformation properties as $k_xk_y$, this places the second constraint on possible allowed little co-groups (the group of crystal symmetries that keep the relevant momentum point unchanged up to a reciprocal lattice vector). Specifically,  the little co-group must have a specific irreducible representation that transforms as $k_xk_y$. This representation should not mix in other quadratic terms, such as $k_x^2$, because that would lead to terms like $k_x^2\tau_x$ in the kp Hamiltonian. Notably, linear terms like $k_x\tau_x$ are already excluded from the kp Hamiltonian by $T$.

Now suppose $\tau_{x,z}$ already transform as $k_xk_y$. Since $\tau_y$ is proportional to the product of $\tau_{x}$ and $\tau_{z}$, it must transform trivially under all crystal symmetries, and belong to the trivial representation. The trivial representation should not mix in linear terms like $k_x$, because that would lead to terms like $k_x\tau_y$ in the Hamiltonian. 


Let us check the above requirements of little co-groups on $\kappa$-Cl, which has mirror symmetries $\widetilde{M_x}=\{M_x|1/2,1/2\}$, $\widetilde{M_y}=\{M_y|1/2,1/2\}$, and their product $C_{2z}$. These three symmetries keep the TRIM point $(\pi,\pi)$ invariant. At this TRIM point, these symmetries form the 2mm point group. In this point group, the $A_2$ representation only hosts $k_xk_y$ at quadratic level. The trivial $A_1$ representation does not contain k-odd terms in $(k_x,k_y)$. The above requirement of little co-groups are thus satisfied.

The third requirement is on the symmetry operators. Because $T$ commutes with all crystal symmetries, these symmetries are represented by { real} $2\times2$ matrices. Since $\tau_y$ is invariant under all crystal symmetries, these matrices should commute with $\tau_y$. Hence, they are either $\pm\tau_0$(identity) or $\pm i\tau_y$. This implies that all the symmetry operators at the TRIM point should commute with each other.

If a symmetry preserves $\tau_{x,z}\sim k_xk_y$, then $\tau_{x,z}$ should commute with the symmetry matrix. The symmetry operator is then $\pm\tau_0$. Such symmetries are thus squared to $+1$. If a symmetry flips $k_xk_y$, then $\tau_{x,z}$ should anticommute with the symmetry matrix. The symmetry matrix is then $\pm i\tau_y$. Such symmetries are thus squared to -1. 

We can now deduce the remaining symmetry requirements for the kp Hamiltonian in Eq.\ref{Eq:kp}: (1) All the symmetry operators at the TRIM point commute with each other. (2) Symmetries preserving $k_xk_y$ are squared to $+1$ while symmetries flipping $k_xk_y$ are squared to $-1$. As shown below, these requirements can be checked without explicitly introducing matrix representations for the symmetries.

Let us check the above requirements for the symmetries on $\kappa$-Cl. Firstly, $\widetilde{M_x}$ and $\widetilde{M_y}$ commute as:
\begin{equation}
\begin{split}
&\{M_x|t^x_x,t^y_x\}\{M_y|t^x_y,t^y_y\}\\=&\{-2t^x_y,2t^y_x\}\{M_y|t^x_y,t^y_y\}\{M_x|t^x_x,t^y_x\}\\=&\{-1,1\}\{M_y|t^x_y,t^y_y\}\{M_x|t^x_x,t^y_x\}
\end{split}
\end{equation}
In real space, the commutator is evaluated into a translation operation $\{-1,1\}$, which is equal to $+1$ at the $(k_x,k_y)=(\pi,\pi)$ point. Symmetries $\widetilde{M_x}$, $\widetilde{M_y}$, and their product $C_{2z}$ thus commute with each other.

Note that $k_xk_y$ is odd under $\widetilde{M_x}$ and $\widetilde{M_y}$, but even under their product $C_{2z}$. Consequently, we require $\widetilde{M_x}^2=\widetilde{M_y}^2=-1$, and $C_{2z}^2=+1$. To check this: 
\begin{equation}
\begin{split}
&\widetilde{M_x}^2=\{0,1\}=\exp(ik_y)=-1\\
&\widetilde{M_y}^2=\{1,0\}=\exp(ik_x)=-1\\
&C_{2z}^2=\{0,0\}=1.
\end{split}
\end{equation}
The symmetry conditions are thus satisfied in $\kappa$-Cl. Table~\ref{T:SG} is obtained by checking the above three requirements for TRIM points in all 2D layer groups and 3D space groups.
}

{\it Discussion:} Here we have identified coincident Van Hove singularities as a platform to realize 2D altermagnetic states {\red and other novel electronic states}. These states are stabilized due to a new interaction term, $g_5$, through a weak-coupling BCS mechanism that leads to altermagnetism, nematic, d-wave superconducting, and orbital altermagnetic orders. Our results apply to {\red nine} 2D space groups and we have chosen a specific application in $\kappa$-Cl, where altermagnetism with the same symmetry found here is observed in $\kappa$-Cl \cite{miyagawa:1995} and d-wave superconductivity is reported under pressure \cite{williams:1990,miyagawa:2004} and anion substitution \cite{kurosaki:2005,miyagawa:2002,kanoda:1997,mayaffre:1995,de:1995,kanoda:1996,urayama:1988,kini:1990}. {\red Another relevant 2D material is monolayer RuF$_4$ in layer group L17(p2$_1$/b11). It has the same kp Hamiltonian as $\kappa$-Cl at the $(\pi,\pi)$ point, and recent DFT calculation \cite{milivojevic:2024} reveals an altermagnetic instability within an RPA-like approach.
}

{\red Although our study focuses on 2D systems, the same coincident Van Hove singularities exist at high symmetry momenta in the 3D space groups listed in Table.\ref{T:SG}. Strictly speaking, in 3D, there is no divergence in the density of states near the Van Hove singularity. However, the density of states can still be large near the Van Hove momentum in the quasi-2D limit. In this case, the BCS weak-coupling instability will still aid in stabilizing the altermagnetic states. Among these 3D space groups, 55(Pbam), 58(Pnnm), 62(Pnma), 136(P$4_2$/mnm), and 138(P$4_2$/ncm) are known to host multiple altermagnetic candidates, according to the search on MAGNDATA database\cite{guo:2023S}. In these 3D materials, the contribution of the coincident Van Hove singularity to the ordering instability can be revealed by studying the doping dependence of the altermagnetic transition temperature.
}

{\it Data Availability:} All study data are included in
the article and/or SI Appendix.

{\it Code Availability:} Codes are available at https://doi.org/10.5281/zenodo.10994906.

\bibliography{citation}

\begin{thebibliography}{88}
\expandafter\ifx\csname natexlab\endcsname\relax\def\natexlab#1{#1}\fi
\expandafter\ifx\csname bibnamefont\endcsname\relax
  \def\bibnamefont#1{#1}\fi
\expandafter\ifx\csname bibfnamefont\endcsname\relax
  \def\bibfnamefont#1{#1}\fi
\expandafter\ifx\csname citenamefont\endcsname\relax
  \def\citenamefont#1{#1}\fi
\expandafter\ifx\csname url\endcsname\relax
  \def\url#1{\texttt{#1}}\fi
\expandafter\ifx\csname urlprefix\endcsname\relax\def\urlprefix{URL }\fi
\providecommand{\bibinfo}[2]{#2}
\providecommand{\eprint}[2][]{\url{#2}}

\bibitem[{\citenamefont{Naka et~al.}(2019)\citenamefont{Naka, Hayami, Kusunose,
  Yanagi, Motome, and Seo}}]{naka:2019}
\bibinfo{author}{\bibfnamefont{M.}~\bibnamefont{Naka}},
  \bibinfo{author}{\bibfnamefont{S.}~\bibnamefont{Hayami}},
  \bibinfo{author}{\bibfnamefont{H.}~\bibnamefont{Kusunose}},
  \bibinfo{author}{\bibfnamefont{Y.}~\bibnamefont{Yanagi}},
  \bibinfo{author}{\bibfnamefont{Y.}~\bibnamefont{Motome}}, \bibnamefont{and}
  \bibinfo{author}{\bibfnamefont{H.}~\bibnamefont{Seo}},
  \bibinfo{journal}{Nature communications} \textbf{\bibinfo{volume}{10}},
  \bibinfo{pages}{4305} (\bibinfo{year}{2019}).

\bibitem[{\citenamefont{{\v{S}}mejkal et~al.}(2020)\citenamefont{{\v{S}}mejkal,
  Gonz{\'a}lez-Hern{\'a}ndez, Jungwirth, and Sinova}}]{vsmejkal:2020}
\bibinfo{author}{\bibfnamefont{L.}~\bibnamefont{{\v{S}}mejkal}},
  \bibinfo{author}{\bibfnamefont{R.}~\bibnamefont{Gonz{\'a}lez-Hern{\'a}ndez}},
  \bibinfo{author}{\bibfnamefont{T.}~\bibnamefont{Jungwirth}},
  \bibnamefont{and} \bibinfo{author}{\bibfnamefont{J.}~\bibnamefont{Sinova}},
  \bibinfo{journal}{Science advances} \textbf{\bibinfo{volume}{6}},
  \bibinfo{pages}{eaaz8809} (\bibinfo{year}{2020}).

\bibitem[{\citenamefont{{\v{S}}mejkal
  et~al.}(2022{\natexlab{a}})\citenamefont{{\v{S}}mejkal, Sinova, and
  Jungwirth}}]{vsmejkal:2022emerging}
\bibinfo{author}{\bibfnamefont{L.}~\bibnamefont{{\v{S}}mejkal}},
  \bibinfo{author}{\bibfnamefont{J.}~\bibnamefont{Sinova}}, \bibnamefont{and}
  \bibinfo{author}{\bibfnamefont{T.}~\bibnamefont{Jungwirth}},
  \bibinfo{journal}{Physical Review X} \textbf{\bibinfo{volume}{12}},
  \bibinfo{pages}{040501} (\bibinfo{year}{2022}{\natexlab{a}}).

\bibitem[{\citenamefont{{\v{S}}mejkal
  et~al.}(2022{\natexlab{b}})\citenamefont{{\v{S}}mejkal, Sinova, and
  Jungwirth}}]{vsmejkal:2022beyond}
\bibinfo{author}{\bibfnamefont{L.}~\bibnamefont{{\v{S}}mejkal}},
  \bibinfo{author}{\bibfnamefont{J.}~\bibnamefont{Sinova}}, \bibnamefont{and}
  \bibinfo{author}{\bibfnamefont{T.}~\bibnamefont{Jungwirth}},
  \bibinfo{journal}{Physical Review X} \textbf{\bibinfo{volume}{12}},
  \bibinfo{pages}{031042} (\bibinfo{year}{2022}{\natexlab{b}}).

\bibitem[{\citenamefont{Noda et~al.}(2016)\citenamefont{Noda, Ohno, and
  Nakamura}}]{noda:2016}
\bibinfo{author}{\bibfnamefont{Y.}~\bibnamefont{Noda}},
  \bibinfo{author}{\bibfnamefont{K.}~\bibnamefont{Ohno}}, \bibnamefont{and}
  \bibinfo{author}{\bibfnamefont{S.}~\bibnamefont{Nakamura}},
  \bibinfo{journal}{Physical Chemistry Chemical Physics}
  \textbf{\bibinfo{volume}{18}}, \bibinfo{pages}{13294} (\bibinfo{year}{2016}).

\bibitem[{\citenamefont{Ahn et~al.}(2019)\citenamefont{Ahn, Hariki, Lee, and
  Kune{\v{s}}}}]{ahn:2019}
\bibinfo{author}{\bibfnamefont{K.-H.} \bibnamefont{Ahn}},
  \bibinfo{author}{\bibfnamefont{A.}~\bibnamefont{Hariki}},
  \bibinfo{author}{\bibfnamefont{K.-W.} \bibnamefont{Lee}}, \bibnamefont{and}
  \bibinfo{author}{\bibfnamefont{J.}~\bibnamefont{Kune{\v{s}}}},
  \bibinfo{journal}{Physical Review B} \textbf{\bibinfo{volume}{99}},
  \bibinfo{pages}{184432} (\bibinfo{year}{2019}).

\bibitem[{\citenamefont{Hayami et~al.}(2019)\citenamefont{Hayami, Yanagi, and
  Kusunose}}]{hayami:2019}
\bibinfo{author}{\bibfnamefont{S.}~\bibnamefont{Hayami}},
  \bibinfo{author}{\bibfnamefont{Y.}~\bibnamefont{Yanagi}}, \bibnamefont{and}
  \bibinfo{author}{\bibfnamefont{H.}~\bibnamefont{Kusunose}},
  \bibinfo{journal}{journal of the physical society of japan}
  \textbf{\bibinfo{volume}{88}}, \bibinfo{pages}{123702}
  (\bibinfo{year}{2019}).

\bibitem[{\citenamefont{Hayami et~al.}(2020)\citenamefont{Hayami, Yanagi, and
  Kusunose}}]{hayami:2020}
\bibinfo{author}{\bibfnamefont{S.}~\bibnamefont{Hayami}},
  \bibinfo{author}{\bibfnamefont{Y.}~\bibnamefont{Yanagi}}, \bibnamefont{and}
  \bibinfo{author}{\bibfnamefont{H.}~\bibnamefont{Kusunose}},
  \bibinfo{journal}{Physical Review B} \textbf{\bibinfo{volume}{102}},
  \bibinfo{pages}{144441} (\bibinfo{year}{2020}).

\bibitem[{\citenamefont{Yuan et~al.}(2021)\citenamefont{Yuan, Wang, Luo, and
  Zunger}}]{yuan:2021}
\bibinfo{author}{\bibfnamefont{L.-D.} \bibnamefont{Yuan}},
  \bibinfo{author}{\bibfnamefont{Z.}~\bibnamefont{Wang}},
  \bibinfo{author}{\bibfnamefont{J.-W.} \bibnamefont{Luo}}, \bibnamefont{and}
  \bibinfo{author}{\bibfnamefont{A.}~\bibnamefont{Zunger}},
  \bibinfo{journal}{Physical Review Materials} \textbf{\bibinfo{volume}{5}},
  \bibinfo{pages}{014409} (\bibinfo{year}{2021}).

\bibitem[{\citenamefont{Mazin et~al.}(2021)\citenamefont{Mazin, Koepernik,
  Johannes, Gonz{\'a}lez-Hern{\'a}ndez, and {\v{S}}mejkal}}]{mazin:2021}
\bibinfo{author}{\bibfnamefont{I.~I.} \bibnamefont{Mazin}},
  \bibinfo{author}{\bibfnamefont{K.}~\bibnamefont{Koepernik}},
  \bibinfo{author}{\bibfnamefont{M.~D.} \bibnamefont{Johannes}},
  \bibinfo{author}{\bibfnamefont{R.}~\bibnamefont{Gonz{\'a}lez-Hern{\'a}ndez}},
  \bibnamefont{and}
  \bibinfo{author}{\bibfnamefont{L.}~\bibnamefont{{\v{S}}mejkal}},
  \bibinfo{journal}{Proceedings of the National Academy of Sciences}
  \textbf{\bibinfo{volume}{118}}, \bibinfo{pages}{e2108924118}
  (\bibinfo{year}{2021}).

\bibitem[{\citenamefont{Berlijn et~al.}(2017)\citenamefont{Berlijn, Snijders,
  Delaire, Zhou, Maier, Cao, Chi, Matsuda, Wang, Koehler
  et~al.}}]{berlijn:2017}
\bibinfo{author}{\bibfnamefont{T.}~\bibnamefont{Berlijn}},
  \bibinfo{author}{\bibfnamefont{P.~C.} \bibnamefont{Snijders}},
  \bibinfo{author}{\bibfnamefont{O.}~\bibnamefont{Delaire}},
  \bibinfo{author}{\bibfnamefont{H.-D.} \bibnamefont{Zhou}},
  \bibinfo{author}{\bibfnamefont{T.~A.} \bibnamefont{Maier}},
  \bibinfo{author}{\bibfnamefont{H.-B.} \bibnamefont{Cao}},
  \bibinfo{author}{\bibfnamefont{S.-X.} \bibnamefont{Chi}},
  \bibinfo{author}{\bibfnamefont{M.}~\bibnamefont{Matsuda}},
  \bibinfo{author}{\bibfnamefont{Y.}~\bibnamefont{Wang}},
  \bibinfo{author}{\bibfnamefont{M.~R.} \bibnamefont{Koehler}},
  \bibnamefont{et~al.}, \bibinfo{journal}{Physical review letters}
  \textbf{\bibinfo{volume}{118}}, \bibinfo{pages}{077201}
  (\bibinfo{year}{2017}).

\bibitem[{\citenamefont{Zhu et~al.}(2019)\citenamefont{Zhu, Strempfer, Rao,
  Occhialini, Pelliciari, Choi, Kawaguchi, You, Mitchell, Shao-Horn
  et~al.}}]{zhu:2019}
\bibinfo{author}{\bibfnamefont{Z.}~\bibnamefont{Zhu}},
  \bibinfo{author}{\bibfnamefont{J.}~\bibnamefont{Strempfer}},
  \bibinfo{author}{\bibfnamefont{R.}~\bibnamefont{Rao}},
  \bibinfo{author}{\bibfnamefont{C.}~\bibnamefont{Occhialini}},
  \bibinfo{author}{\bibfnamefont{J.}~\bibnamefont{Pelliciari}},
  \bibinfo{author}{\bibfnamefont{Y.}~\bibnamefont{Choi}},
  \bibinfo{author}{\bibfnamefont{T.}~\bibnamefont{Kawaguchi}},
  \bibinfo{author}{\bibfnamefont{H.}~\bibnamefont{You}},
  \bibinfo{author}{\bibfnamefont{J.}~\bibnamefont{Mitchell}},
  \bibinfo{author}{\bibfnamefont{Y.}~\bibnamefont{Shao-Horn}},
  \bibnamefont{et~al.}, \bibinfo{journal}{Physical review letters}
  \textbf{\bibinfo{volume}{122}}, \bibinfo{pages}{017202}
  (\bibinfo{year}{2019}).

\bibitem[{\citenamefont{Kivelson et~al.}(2003)\citenamefont{Kivelson, Bindloss,
  Fradkin, Oganesyan, Tranquada, Kapitulnik, and Howald}}]{kivelson:2003}
\bibinfo{author}{\bibfnamefont{S.~A.} \bibnamefont{Kivelson}},
  \bibinfo{author}{\bibfnamefont{I.~P.} \bibnamefont{Bindloss}},
  \bibinfo{author}{\bibfnamefont{E.}~\bibnamefont{Fradkin}},
  \bibinfo{author}{\bibfnamefont{V.}~\bibnamefont{Oganesyan}},
  \bibinfo{author}{\bibfnamefont{J.}~\bibnamefont{Tranquada}},
  \bibinfo{author}{\bibfnamefont{A.}~\bibnamefont{Kapitulnik}},
  \bibnamefont{and} \bibinfo{author}{\bibfnamefont{C.}~\bibnamefont{Howald}},
  \bibinfo{journal}{Reviews of Modern Physics} \textbf{\bibinfo{volume}{75}},
  \bibinfo{pages}{1201} (\bibinfo{year}{2003}).

\bibitem[{\citenamefont{{\v{S}}mejkal
  et~al.}(2022{\natexlab{c}})\citenamefont{{\v{S}}mejkal, MacDonald, Sinova,
  Nakatsuji, and Jungwirth}}]{vsmejkal:2022}
\bibinfo{author}{\bibfnamefont{L.}~\bibnamefont{{\v{S}}mejkal}},
  \bibinfo{author}{\bibfnamefont{A.~H.} \bibnamefont{MacDonald}},
  \bibinfo{author}{\bibfnamefont{J.}~\bibnamefont{Sinova}},
  \bibinfo{author}{\bibfnamefont{S.}~\bibnamefont{Nakatsuji}},
  \bibnamefont{and}
  \bibinfo{author}{\bibfnamefont{T.}~\bibnamefont{Jungwirth}},
  \bibinfo{journal}{Nature Reviews Materials} \textbf{\bibinfo{volume}{7}},
  \bibinfo{pages}{482} (\bibinfo{year}{2022}{\natexlab{c}}).

\bibitem[{\citenamefont{Feng et~al.}(2022)\citenamefont{Feng, Zhou,
  {\v{S}}mejkal, Wu, Zhu, Guo, Gonz{\'a}lez-Hern{\'a}ndez, Wang, Yan, Qin
  et~al.}}]{feng:2022}
\bibinfo{author}{\bibfnamefont{Z.}~\bibnamefont{Feng}},
  \bibinfo{author}{\bibfnamefont{X.}~\bibnamefont{Zhou}},
  \bibinfo{author}{\bibfnamefont{L.}~\bibnamefont{{\v{S}}mejkal}},
  \bibinfo{author}{\bibfnamefont{L.}~\bibnamefont{Wu}},
  \bibinfo{author}{\bibfnamefont{Z.}~\bibnamefont{Zhu}},
  \bibinfo{author}{\bibfnamefont{H.}~\bibnamefont{Guo}},
  \bibinfo{author}{\bibfnamefont{R.}~\bibnamefont{Gonz{\'a}lez-Hern{\'a}ndez}},
  \bibinfo{author}{\bibfnamefont{X.}~\bibnamefont{Wang}},
  \bibinfo{author}{\bibfnamefont{H.}~\bibnamefont{Yan}},
  \bibinfo{author}{\bibfnamefont{P.}~\bibnamefont{Qin}}, \bibnamefont{et~al.},
  \bibinfo{journal}{Nature Electronics} \textbf{\bibinfo{volume}{5}},
  \bibinfo{pages}{735} (\bibinfo{year}{2022}).

\bibitem[{\citenamefont{Betancourt et~al.}(2023)\citenamefont{Betancourt,
  Zub{\'a}{\v{c}}, Gonzalez-Hernandez, Geishendorf, {\v{S}}ob{\'a}{\v{n}},
  Springholz, Olejn{\'\i}k, {\v{S}}mejkal, Sinova, Jungwirth
  et~al.}}]{betancourt:2023}
\bibinfo{author}{\bibfnamefont{R.~G.} \bibnamefont{Betancourt}},
  \bibinfo{author}{\bibfnamefont{J.}~\bibnamefont{Zub{\'a}{\v{c}}}},
  \bibinfo{author}{\bibfnamefont{R.}~\bibnamefont{Gonzalez-Hernandez}},
  \bibinfo{author}{\bibfnamefont{K.}~\bibnamefont{Geishendorf}},
  \bibinfo{author}{\bibfnamefont{Z.}~\bibnamefont{{\v{S}}ob{\'a}{\v{n}}}},
  \bibinfo{author}{\bibfnamefont{G.}~\bibnamefont{Springholz}},
  \bibinfo{author}{\bibfnamefont{K.}~\bibnamefont{Olejn{\'\i}k}},
  \bibinfo{author}{\bibfnamefont{L.}~\bibnamefont{{\v{S}}mejkal}},
  \bibinfo{author}{\bibfnamefont{J.}~\bibnamefont{Sinova}},
  \bibinfo{author}{\bibfnamefont{T.}~\bibnamefont{Jungwirth}},
  \bibnamefont{et~al.}, \bibinfo{journal}{Physical Review Letters}
  \textbf{\bibinfo{volume}{130}}, \bibinfo{pages}{036702}
  (\bibinfo{year}{2023}).

\bibitem[{\citenamefont{Naka et~al.}(2020)\citenamefont{Naka, Hayami, Kusunose,
  Yanagi, Motome, and Seo}}]{naka:2020}
\bibinfo{author}{\bibfnamefont{M.}~\bibnamefont{Naka}},
  \bibinfo{author}{\bibfnamefont{S.}~\bibnamefont{Hayami}},
  \bibinfo{author}{\bibfnamefont{H.}~\bibnamefont{Kusunose}},
  \bibinfo{author}{\bibfnamefont{Y.}~\bibnamefont{Yanagi}},
  \bibinfo{author}{\bibfnamefont{Y.}~\bibnamefont{Motome}}, \bibnamefont{and}
  \bibinfo{author}{\bibfnamefont{H.}~\bibnamefont{Seo}},
  \bibinfo{journal}{Physical Review B} \textbf{\bibinfo{volume}{102}},
  \bibinfo{pages}{075112} (\bibinfo{year}{2020}).

\bibitem[{\citenamefont{Nakatsuji et~al.}(2015)\citenamefont{Nakatsuji,
  Kiyohara, and Higo}}]{nakatsuji:2015}
\bibinfo{author}{\bibfnamefont{S.}~\bibnamefont{Nakatsuji}},
  \bibinfo{author}{\bibfnamefont{N.}~\bibnamefont{Kiyohara}}, \bibnamefont{and}
  \bibinfo{author}{\bibfnamefont{T.}~\bibnamefont{Higo}},
  \bibinfo{journal}{Nature} \textbf{\bibinfo{volume}{527}},
  \bibinfo{pages}{212} (\bibinfo{year}{2015}).

\bibitem[{\citenamefont{Samanta et~al.}(2020)\citenamefont{Samanta,
  Le{\v{z}}ai{\'c}, Merte, Freimuth, Bl{\"u}gel, and Mokrousov}}]{samanta:2020}
\bibinfo{author}{\bibfnamefont{K.}~\bibnamefont{Samanta}},
  \bibinfo{author}{\bibfnamefont{M.}~\bibnamefont{Le{\v{z}}ai{\'c}}},
  \bibinfo{author}{\bibfnamefont{M.}~\bibnamefont{Merte}},
  \bibinfo{author}{\bibfnamefont{F.}~\bibnamefont{Freimuth}},
  \bibinfo{author}{\bibfnamefont{S.}~\bibnamefont{Bl{\"u}gel}},
  \bibnamefont{and}
  \bibinfo{author}{\bibfnamefont{Y.}~\bibnamefont{Mokrousov}},
  \bibinfo{journal}{Journal of applied physics} \textbf{\bibinfo{volume}{127}}
  (\bibinfo{year}{2020}).

\bibitem[{\citenamefont{S{\"u}rgers et~al.}(2016)\citenamefont{S{\"u}rgers,
  Kittler, Wolf, and L{\"o}hneysen}}]{surgers:2016}
\bibinfo{author}{\bibfnamefont{C.}~\bibnamefont{S{\"u}rgers}},
  \bibinfo{author}{\bibfnamefont{W.}~\bibnamefont{Kittler}},
  \bibinfo{author}{\bibfnamefont{T.}~\bibnamefont{Wolf}}, \bibnamefont{and}
  \bibinfo{author}{\bibfnamefont{H.~v.} \bibnamefont{L{\"o}hneysen}},
  \bibinfo{journal}{AIP Advances} \textbf{\bibinfo{volume}{6}}
  (\bibinfo{year}{2016}).

\bibitem[{\citenamefont{Ghimire et~al.}(2018)\citenamefont{Ghimire, Botana,
  Jiang, Zhang, Chen, and Mitchell}}]{ghimire:2018}
\bibinfo{author}{\bibfnamefont{N.~J.} \bibnamefont{Ghimire}},
  \bibinfo{author}{\bibfnamefont{A.}~\bibnamefont{Botana}},
  \bibinfo{author}{\bibfnamefont{J.}~\bibnamefont{Jiang}},
  \bibinfo{author}{\bibfnamefont{J.}~\bibnamefont{Zhang}},
  \bibinfo{author}{\bibfnamefont{Y.-S.} \bibnamefont{Chen}}, \bibnamefont{and}
  \bibinfo{author}{\bibfnamefont{J.}~\bibnamefont{Mitchell}},
  \bibinfo{journal}{Nature communications} \textbf{\bibinfo{volume}{9}},
  \bibinfo{pages}{3280} (\bibinfo{year}{2018}).

\bibitem[{\citenamefont{Jungwirth et~al.}(2016)\citenamefont{Jungwirth, Marti,
  Wadley, and Wunderlich}}]{jungwirth:2016}
\bibinfo{author}{\bibfnamefont{T.}~\bibnamefont{Jungwirth}},
  \bibinfo{author}{\bibfnamefont{X.}~\bibnamefont{Marti}},
  \bibinfo{author}{\bibfnamefont{P.}~\bibnamefont{Wadley}}, \bibnamefont{and}
  \bibinfo{author}{\bibfnamefont{J.}~\bibnamefont{Wunderlich}},
  \bibinfo{journal}{Nature nanotechnology} \textbf{\bibinfo{volume}{11}},
  \bibinfo{pages}{231} (\bibinfo{year}{2016}).

\bibitem[{\citenamefont{{\v{S}}mejkal
  et~al.}(2022{\natexlab{d}})\citenamefont{{\v{S}}mejkal, Hellenes,
  Gonz{\'a}lez-Hern{\'a}ndez, Sinova, and Jungwirth}}]{vsmejkal:2022giant}
\bibinfo{author}{\bibfnamefont{L.}~\bibnamefont{{\v{S}}mejkal}},
  \bibinfo{author}{\bibfnamefont{A.~B.} \bibnamefont{Hellenes}},
  \bibinfo{author}{\bibfnamefont{R.}~\bibnamefont{Gonz{\'a}lez-Hern{\'a}ndez}},
  \bibinfo{author}{\bibfnamefont{J.}~\bibnamefont{Sinova}}, \bibnamefont{and}
  \bibinfo{author}{\bibfnamefont{T.}~\bibnamefont{Jungwirth}},
  \bibinfo{journal}{Physical Review X} \textbf{\bibinfo{volume}{12}},
  \bibinfo{pages}{011028} (\bibinfo{year}{2022}{\natexlab{d}}).

\bibitem[{\citenamefont{Hariki et~al.}(2023)\citenamefont{Hariki, Yamaguchi,
  Kriegner, Edmonds, Wadley, Dhesi, Springholz, {\v{S}}mejkal, V{\`y}born{\`y},
  Jungwirth et~al.}}]{hariki:2023}
\bibinfo{author}{\bibfnamefont{A.}~\bibnamefont{Hariki}},
  \bibinfo{author}{\bibfnamefont{T.}~\bibnamefont{Yamaguchi}},
  \bibinfo{author}{\bibfnamefont{D.}~\bibnamefont{Kriegner}},
  \bibinfo{author}{\bibfnamefont{K.}~\bibnamefont{Edmonds}},
  \bibinfo{author}{\bibfnamefont{P.}~\bibnamefont{Wadley}},
  \bibinfo{author}{\bibfnamefont{S.}~\bibnamefont{Dhesi}},
  \bibinfo{author}{\bibfnamefont{G.}~\bibnamefont{Springholz}},
  \bibinfo{author}{\bibfnamefont{L.}~\bibnamefont{{\v{S}}mejkal}},
  \bibinfo{author}{\bibfnamefont{K.}~\bibnamefont{V{\`y}born{\`y}}},
  \bibinfo{author}{\bibfnamefont{T.}~\bibnamefont{Jungwirth}},
  \bibnamefont{et~al.}, \bibinfo{journal}{arXiv preprint arXiv:2305.03588}
  (\bibinfo{year}{2023}).

\bibitem[{\citenamefont{Yuan et~al.}(2020)\citenamefont{Yuan, Wang, Luo,
  Rashba, and Zunger}}]{yuan:2020}
\bibinfo{author}{\bibfnamefont{L.-D.} \bibnamefont{Yuan}},
  \bibinfo{author}{\bibfnamefont{Z.}~\bibnamefont{Wang}},
  \bibinfo{author}{\bibfnamefont{J.-W.} \bibnamefont{Luo}},
  \bibinfo{author}{\bibfnamefont{E.~I.} \bibnamefont{Rashba}},
  \bibnamefont{and} \bibinfo{author}{\bibfnamefont{A.}~\bibnamefont{Zunger}},
  \bibinfo{journal}{Physical Review B} \textbf{\bibinfo{volume}{102}},
  \bibinfo{pages}{014422} (\bibinfo{year}{2020}).

\bibitem[{\citenamefont{Gonz{\'a}lez-Hern{\'a}ndez
  et~al.}(2021)\citenamefont{Gonz{\'a}lez-Hern{\'a}ndez, {\v{S}}mejkal,
  V{\`y}born{\`y}, Yahagi, Sinova, Jungwirth, and
  {\v{Z}}elezn{\`y}}}]{gonzalez:2021}
\bibinfo{author}{\bibfnamefont{R.}~\bibnamefont{Gonz{\'a}lez-Hern{\'a}ndez}},
  \bibinfo{author}{\bibfnamefont{L.}~\bibnamefont{{\v{S}}mejkal}},
  \bibinfo{author}{\bibfnamefont{K.}~\bibnamefont{V{\`y}born{\`y}}},
  \bibinfo{author}{\bibfnamefont{Y.}~\bibnamefont{Yahagi}},
  \bibinfo{author}{\bibfnamefont{J.}~\bibnamefont{Sinova}},
  \bibinfo{author}{\bibfnamefont{T.}~\bibnamefont{Jungwirth}},
  \bibnamefont{and}
  \bibinfo{author}{\bibfnamefont{J.}~\bibnamefont{{\v{Z}}elezn{\`y}}},
  \bibinfo{journal}{Physical Review Letters} \textbf{\bibinfo{volume}{126}},
  \bibinfo{pages}{127701} (\bibinfo{year}{2021}).

\bibitem[{\citenamefont{Naka et~al.}(2021)\citenamefont{Naka, Motome, and
  Seo}}]{naka:2021}
\bibinfo{author}{\bibfnamefont{M.}~\bibnamefont{Naka}},
  \bibinfo{author}{\bibfnamefont{Y.}~\bibnamefont{Motome}}, \bibnamefont{and}
  \bibinfo{author}{\bibfnamefont{H.}~\bibnamefont{Seo}},
  \bibinfo{journal}{Physical Review B} \textbf{\bibinfo{volume}{103}},
  \bibinfo{pages}{125114} (\bibinfo{year}{2021}).

\bibitem[{\citenamefont{Shao et~al.}(2021)\citenamefont{Shao, Zhang, Li, Eom,
  and Tsymbal}}]{shao:2021}
\bibinfo{author}{\bibfnamefont{D.-F.} \bibnamefont{Shao}},
  \bibinfo{author}{\bibfnamefont{S.-H.} \bibnamefont{Zhang}},
  \bibinfo{author}{\bibfnamefont{M.}~\bibnamefont{Li}},
  \bibinfo{author}{\bibfnamefont{C.-B.} \bibnamefont{Eom}}, \bibnamefont{and}
  \bibinfo{author}{\bibfnamefont{E.~Y.} \bibnamefont{Tsymbal}},
  \bibinfo{journal}{Nature Communications} \textbf{\bibinfo{volume}{12}},
  \bibinfo{pages}{7061} (\bibinfo{year}{2021}).

\bibitem[{\citenamefont{Bose et~al.}(2022)\citenamefont{Bose, Schreiber, Jain,
  Shao, Nair, Sun, Zhang, Muller, Tsymbal, Schlom et~al.}}]{bose:2022}
\bibinfo{author}{\bibfnamefont{A.}~\bibnamefont{Bose}},
  \bibinfo{author}{\bibfnamefont{N.~J.} \bibnamefont{Schreiber}},
  \bibinfo{author}{\bibfnamefont{R.}~\bibnamefont{Jain}},
  \bibinfo{author}{\bibfnamefont{D.-F.} \bibnamefont{Shao}},
  \bibinfo{author}{\bibfnamefont{H.~P.} \bibnamefont{Nair}},
  \bibinfo{author}{\bibfnamefont{J.}~\bibnamefont{Sun}},
  \bibinfo{author}{\bibfnamefont{X.~S.} \bibnamefont{Zhang}},
  \bibinfo{author}{\bibfnamefont{D.~A.} \bibnamefont{Muller}},
  \bibinfo{author}{\bibfnamefont{E.~Y.} \bibnamefont{Tsymbal}},
  \bibinfo{author}{\bibfnamefont{D.~G.} \bibnamefont{Schlom}},
  \bibnamefont{et~al.}, \bibinfo{journal}{Nature Electronics}
  \textbf{\bibinfo{volume}{5}}, \bibinfo{pages}{267} (\bibinfo{year}{2022}).

\bibitem[{\citenamefont{Ma et~al.}(2021)\citenamefont{Ma, Hu, Li, Liu, Yao,
  Jia, and Liu}}]{ma:2021}
\bibinfo{author}{\bibfnamefont{H.-Y.} \bibnamefont{Ma}},
  \bibinfo{author}{\bibfnamefont{M.}~\bibnamefont{Hu}},
  \bibinfo{author}{\bibfnamefont{N.}~\bibnamefont{Li}},
  \bibinfo{author}{\bibfnamefont{J.}~\bibnamefont{Liu}},
  \bibinfo{author}{\bibfnamefont{W.}~\bibnamefont{Yao}},
  \bibinfo{author}{\bibfnamefont{J.-F.} \bibnamefont{Jia}}, \bibnamefont{and}
  \bibinfo{author}{\bibfnamefont{J.}~\bibnamefont{Liu}},
  \bibinfo{journal}{Nature communications} \textbf{\bibinfo{volume}{12}},
  \bibinfo{pages}{2846} (\bibinfo{year}{2021}).

\bibitem[{\citenamefont{Guo et~al.}(2023{\natexlab{a}})\citenamefont{Guo, Guo,
  Cheng, Wang, and Ang}}]{guo:2023}
\bibinfo{author}{\bibfnamefont{S.-D.} \bibnamefont{Guo}},
  \bibinfo{author}{\bibfnamefont{X.-S.} \bibnamefont{Guo}},
  \bibinfo{author}{\bibfnamefont{K.}~\bibnamefont{Cheng}},
  \bibinfo{author}{\bibfnamefont{K.}~\bibnamefont{Wang}}, \bibnamefont{and}
  \bibinfo{author}{\bibfnamefont{Y.~S.} \bibnamefont{Ang}},
  \bibinfo{journal}{arXiv preprint arXiv:2306.04094}
  (\bibinfo{year}{2023}{\natexlab{a}}).

\bibitem[{\citenamefont{Steward et~al.}(2023)\citenamefont{Steward, Fernandes,
  and Schmalian}}]{steward:2023}
\bibinfo{author}{\bibfnamefont{C.~R.} \bibnamefont{Steward}},
  \bibinfo{author}{\bibfnamefont{R.~M.} \bibnamefont{Fernandes}},
  \bibnamefont{and}
  \bibinfo{author}{\bibfnamefont{J.}~\bibnamefont{Schmalian}},
  \bibinfo{journal}{Physical Review B} \textbf{\bibinfo{volume}{108}},
  \bibinfo{pages}{144418} (\bibinfo{year}{2023}).

\bibitem[{\citenamefont{Liu et~al.}(2018)\citenamefont{Liu, Chen, Wang, Liu,
  Wang, Feng, Yan, Wang, Jiang, Coey et~al.}}]{liu:2018}
\bibinfo{author}{\bibfnamefont{Z.}~\bibnamefont{Liu}},
  \bibinfo{author}{\bibfnamefont{H.}~\bibnamefont{Chen}},
  \bibinfo{author}{\bibfnamefont{J.}~\bibnamefont{Wang}},
  \bibinfo{author}{\bibfnamefont{J.}~\bibnamefont{Liu}},
  \bibinfo{author}{\bibfnamefont{K.}~\bibnamefont{Wang}},
  \bibinfo{author}{\bibfnamefont{Z.}~\bibnamefont{Feng}},
  \bibinfo{author}{\bibfnamefont{H.}~\bibnamefont{Yan}},
  \bibinfo{author}{\bibfnamefont{X.}~\bibnamefont{Wang}},
  \bibinfo{author}{\bibfnamefont{C.}~\bibnamefont{Jiang}},
  \bibinfo{author}{\bibfnamefont{J.}~\bibnamefont{Coey}}, \bibnamefont{et~al.},
  \bibinfo{journal}{Nature Electronics} \textbf{\bibinfo{volume}{1}},
  \bibinfo{pages}{172} (\bibinfo{year}{2018}).

\bibitem[{\citenamefont{L{\'o}pez-Moreno
  et~al.}(2012)\citenamefont{L{\'o}pez-Moreno, Romero, Mej{\'\i}a-L{\'o}pez,
  Mu{\~n}oz, and Roshchin}}]{lopez:2012}
\bibinfo{author}{\bibfnamefont{S.}~\bibnamefont{L{\'o}pez-Moreno}},
  \bibinfo{author}{\bibfnamefont{A.}~\bibnamefont{Romero}},
  \bibinfo{author}{\bibfnamefont{J.}~\bibnamefont{Mej{\'\i}a-L{\'o}pez}},
  \bibinfo{author}{\bibfnamefont{A.}~\bibnamefont{Mu{\~n}oz}},
  \bibnamefont{and} \bibinfo{author}{\bibfnamefont{I.~V.}
  \bibnamefont{Roshchin}}, \bibinfo{journal}{Physical Review B}
  \textbf{\bibinfo{volume}{85}}, \bibinfo{pages}{134110}
  (\bibinfo{year}{2012}).

\bibitem[{\citenamefont{Karube et~al.}(2022)\citenamefont{Karube, Tanaka,
  Sugawara, Kadoguchi, Kohda, and Nitta}}]{karube:2022}
\bibinfo{author}{\bibfnamefont{S.}~\bibnamefont{Karube}},
  \bibinfo{author}{\bibfnamefont{T.}~\bibnamefont{Tanaka}},
  \bibinfo{author}{\bibfnamefont{D.}~\bibnamefont{Sugawara}},
  \bibinfo{author}{\bibfnamefont{N.}~\bibnamefont{Kadoguchi}},
  \bibinfo{author}{\bibfnamefont{M.}~\bibnamefont{Kohda}}, \bibnamefont{and}
  \bibinfo{author}{\bibfnamefont{J.}~\bibnamefont{Nitta}},
  \bibinfo{journal}{Physical review letters} \textbf{\bibinfo{volume}{129}},
  \bibinfo{pages}{137201} (\bibinfo{year}{2022}).

\bibitem[{\citenamefont{Bai et~al.}(2022)\citenamefont{Bai, Han, Feng, Zhou,
  Su, Wang, Liao, Zhu, Chen, Pan et~al.}}]{bai:2022}
\bibinfo{author}{\bibfnamefont{H.}~\bibnamefont{Bai}},
  \bibinfo{author}{\bibfnamefont{L.}~\bibnamefont{Han}},
  \bibinfo{author}{\bibfnamefont{X.}~\bibnamefont{Feng}},
  \bibinfo{author}{\bibfnamefont{Y.}~\bibnamefont{Zhou}},
  \bibinfo{author}{\bibfnamefont{R.}~\bibnamefont{Su}},
  \bibinfo{author}{\bibfnamefont{Q.}~\bibnamefont{Wang}},
  \bibinfo{author}{\bibfnamefont{L.}~\bibnamefont{Liao}},
  \bibinfo{author}{\bibfnamefont{W.}~\bibnamefont{Zhu}},
  \bibinfo{author}{\bibfnamefont{X.}~\bibnamefont{Chen}},
  \bibinfo{author}{\bibfnamefont{F.}~\bibnamefont{Pan}}, \bibnamefont{et~al.},
  \bibinfo{journal}{Physical Review Letters} \textbf{\bibinfo{volume}{128}},
  \bibinfo{pages}{197202} (\bibinfo{year}{2022}).

\bibitem[{\citenamefont{Tomczak}(2018)}]{tomczak:2018}
\bibinfo{author}{\bibfnamefont{J.~M.} \bibnamefont{Tomczak}},
  \bibinfo{journal}{Journal of Physics: Condensed Matter}
  \textbf{\bibinfo{volume}{30}}, \bibinfo{pages}{183001}
  (\bibinfo{year}{2018}).

\bibitem[{\citenamefont{Zhou et~al.}(2023)\citenamefont{Zhou, Feng, Zhang,
  Smejkal, Sinova, Mokrousov, and Yao}}]{zhou:2023}
\bibinfo{author}{\bibfnamefont{X.}~\bibnamefont{Zhou}},
  \bibinfo{author}{\bibfnamefont{W.}~\bibnamefont{Feng}},
  \bibinfo{author}{\bibfnamefont{R.-W.} \bibnamefont{Zhang}},
  \bibinfo{author}{\bibfnamefont{L.}~\bibnamefont{Smejkal}},
  \bibinfo{author}{\bibfnamefont{J.}~\bibnamefont{Sinova}},
  \bibinfo{author}{\bibfnamefont{Y.}~\bibnamefont{Mokrousov}},
  \bibnamefont{and} \bibinfo{author}{\bibfnamefont{Y.}~\bibnamefont{Yao}},
  \bibinfo{journal}{arXiv preprint arXiv:2305.01410}  (\bibinfo{year}{2023}).

\bibitem[{\citenamefont{Papaj}(2023)}]{papaj:2023}
\bibinfo{author}{\bibfnamefont{M.}~\bibnamefont{Papaj}},
  \bibinfo{journal}{arXiv preprint arXiv:2305.03856}  (\bibinfo{year}{2023}).

\bibitem[{\citenamefont{Beenakker and Vakhtel}(2023)}]{beenakker:2023}
\bibinfo{author}{\bibfnamefont{C.}~\bibnamefont{Beenakker}} \bibnamefont{and}
  \bibinfo{author}{\bibfnamefont{T.}~\bibnamefont{Vakhtel}},
  \bibinfo{journal}{Physical Review B} \textbf{\bibinfo{volume}{108}},
  \bibinfo{pages}{075425} (\bibinfo{year}{2023}).

\bibitem[{\citenamefont{Zhu et~al.}(2023)\citenamefont{Zhu, Zhuang, Wu, and
  Yan}}]{zhu:2023}
\bibinfo{author}{\bibfnamefont{D.}~\bibnamefont{Zhu}},
  \bibinfo{author}{\bibfnamefont{Z.-Y.} \bibnamefont{Zhuang}},
  \bibinfo{author}{\bibfnamefont{Z.}~\bibnamefont{Wu}}, \bibnamefont{and}
  \bibinfo{author}{\bibfnamefont{Z.}~\bibnamefont{Yan}},
  \bibinfo{journal}{arXiv preprint arXiv:2305.10479}  (\bibinfo{year}{2023}).

\bibitem[{\citenamefont{Zhang et~al.}(2023)\citenamefont{Zhang, Hu, and
  Neupert}}]{zhang:2023}
\bibinfo{author}{\bibfnamefont{S.-B.} \bibnamefont{Zhang}},
  \bibinfo{author}{\bibfnamefont{L.-H.} \bibnamefont{Hu}}, \bibnamefont{and}
  \bibinfo{author}{\bibfnamefont{T.}~\bibnamefont{Neupert}},
  \bibinfo{journal}{arXiv preprint arXiv:2302.13185}  (\bibinfo{year}{2023}).

\bibitem[{\citenamefont{Sumita et~al.}(2023)\citenamefont{Sumita, Naka, and
  Seo}}]{sumita:2023}
\bibinfo{author}{\bibfnamefont{S.}~\bibnamefont{Sumita}},
  \bibinfo{author}{\bibfnamefont{M.}~\bibnamefont{Naka}}, \bibnamefont{and}
  \bibinfo{author}{\bibfnamefont{H.}~\bibnamefont{Seo}},
  \bibinfo{journal}{arXiv preprint arXiv:2308.14227}  (\bibinfo{year}{2023}).

\bibitem[{\citenamefont{Ghorashi et~al.}(2023)\citenamefont{Ghorashi, Hughes,
  and Cano}}]{ghorashi:2023}
\bibinfo{author}{\bibfnamefont{S.~A.~A.} \bibnamefont{Ghorashi}},
  \bibinfo{author}{\bibfnamefont{T.~L.} \bibnamefont{Hughes}},
  \bibnamefont{and} \bibinfo{author}{\bibfnamefont{J.}~\bibnamefont{Cano}},
  \bibinfo{journal}{arXiv preprint arXiv:2306.09413}  (\bibinfo{year}{2023}).

\bibitem[{\citenamefont{Roig et~al.}(2024)\citenamefont{Roig, Kreisel, Yu,
  Andersen, and Agterberg}}]{roig:2024}
\bibinfo{author}{\bibfnamefont{M.}~\bibnamefont{Roig}},
  \bibinfo{author}{\bibfnamefont{A.}~\bibnamefont{Kreisel}},
  \bibinfo{author}{\bibfnamefont{Y.}~\bibnamefont{Yu}},
  \bibinfo{author}{\bibfnamefont{B.}~\bibnamefont{Andersen}}, \bibnamefont{and}
  \bibinfo{author}{\bibfnamefont{D.}~\bibnamefont{Agterberg}},
  \bibinfo{journal}{arXiv preprint}  (\bibinfo{year}{2024}).

\bibitem[{\citenamefont{Wu et~al.}(2007)\citenamefont{Wu, Sun, Fradkin, and
  Zhang}}]{Wu:2007}
\bibinfo{author}{\bibfnamefont{C.}~\bibnamefont{Wu}},
  \bibinfo{author}{\bibfnamefont{K.}~\bibnamefont{Sun}},
  \bibinfo{author}{\bibfnamefont{E.}~\bibnamefont{Fradkin}}, \bibnamefont{and}
  \bibinfo{author}{\bibfnamefont{S.-C.} \bibnamefont{Zhang}},
  \bibinfo{journal}{Phys. Rev. B} \textbf{\bibinfo{volume}{75}},
  \bibinfo{pages}{115103} (\bibinfo{year}{2007}),
  \urlprefix\url{https://link.aps.org/doi/10.1103/PhysRevB.75.115103}.

\bibitem[{\citenamefont{Kiesel et~al.}(2012)\citenamefont{Kiesel, Platt, Hanke,
  Abanin, and Thomale}}]{kiesel:2012}
\bibinfo{author}{\bibfnamefont{M.~L.} \bibnamefont{Kiesel}},
  \bibinfo{author}{\bibfnamefont{C.}~\bibnamefont{Platt}},
  \bibinfo{author}{\bibfnamefont{W.}~\bibnamefont{Hanke}},
  \bibinfo{author}{\bibfnamefont{D.~A.} \bibnamefont{Abanin}},
  \bibnamefont{and} \bibinfo{author}{\bibfnamefont{R.}~\bibnamefont{Thomale}},
  \bibinfo{journal}{Physical Review B} \textbf{\bibinfo{volume}{86}},
  \bibinfo{pages}{020507} (\bibinfo{year}{2012}).

\bibitem[{\citenamefont{Nandkishore et~al.}(2012)\citenamefont{Nandkishore,
  Levitov, and Chubukov}}]{nandkishore:2012}
\bibinfo{author}{\bibfnamefont{R.}~\bibnamefont{Nandkishore}},
  \bibinfo{author}{\bibfnamefont{L.~S.} \bibnamefont{Levitov}},
  \bibnamefont{and} \bibinfo{author}{\bibfnamefont{A.~V.}
  \bibnamefont{Chubukov}}, \bibinfo{journal}{Nature Physics}
  \textbf{\bibinfo{volume}{8}}, \bibinfo{pages}{158} (\bibinfo{year}{2012}).

\bibitem[{\citenamefont{Wang et~al.}(2012)\citenamefont{Wang, Xiang, Wang,
  Wang, Yang, and Lee}}]{wang:2012}
\bibinfo{author}{\bibfnamefont{W.-S.} \bibnamefont{Wang}},
  \bibinfo{author}{\bibfnamefont{Y.-Y.} \bibnamefont{Xiang}},
  \bibinfo{author}{\bibfnamefont{Q.-H.} \bibnamefont{Wang}},
  \bibinfo{author}{\bibfnamefont{F.}~\bibnamefont{Wang}},
  \bibinfo{author}{\bibfnamefont{F.}~\bibnamefont{Yang}}, \bibnamefont{and}
  \bibinfo{author}{\bibfnamefont{D.-H.} \bibnamefont{Lee}},
  \bibinfo{journal}{Physical Review B} \textbf{\bibinfo{volume}{85}},
  \bibinfo{pages}{035414} (\bibinfo{year}{2012}).

\bibitem[{\citenamefont{Furukawa et~al.}(1998)\citenamefont{Furukawa, Rice, and
  Salmhofer}}]{furukawa:1998}
\bibinfo{author}{\bibfnamefont{N.}~\bibnamefont{Furukawa}},
  \bibinfo{author}{\bibfnamefont{T.}~\bibnamefont{Rice}}, \bibnamefont{and}
  \bibinfo{author}{\bibfnamefont{M.}~\bibnamefont{Salmhofer}},
  \bibinfo{journal}{Physical review letters} \textbf{\bibinfo{volume}{81}},
  \bibinfo{pages}{3195} (\bibinfo{year}{1998}).

\bibitem[{\citenamefont{Kampf and Katanin}(2003)}]{kampf:2003}
\bibinfo{author}{\bibfnamefont{A.~P.} \bibnamefont{Kampf}} \bibnamefont{and}
  \bibinfo{author}{\bibfnamefont{A.}~\bibnamefont{Katanin}},
  \bibinfo{journal}{Physical Review B} \textbf{\bibinfo{volume}{67}},
  \bibinfo{pages}{125104} (\bibinfo{year}{2003}).

\bibitem[{\citenamefont{Honerkamp et~al.}(2001)\citenamefont{Honerkamp,
  Salmhofer, Furukawa, and Rice}}]{honerkamp:2001}
\bibinfo{author}{\bibfnamefont{C.}~\bibnamefont{Honerkamp}},
  \bibinfo{author}{\bibfnamefont{M.}~\bibnamefont{Salmhofer}},
  \bibinfo{author}{\bibfnamefont{N.}~\bibnamefont{Furukawa}}, \bibnamefont{and}
  \bibinfo{author}{\bibfnamefont{T.~M.} \bibnamefont{Rice}},
  \bibinfo{journal}{Physical Review B} \textbf{\bibinfo{volume}{63}},
  \bibinfo{pages}{035109} (\bibinfo{year}{2001}).

\bibitem[{\citenamefont{Le~Hur and Rice}(2009)}]{le:2009}
\bibinfo{author}{\bibfnamefont{K.}~\bibnamefont{Le~Hur}} \bibnamefont{and}
  \bibinfo{author}{\bibfnamefont{T.~M.} \bibnamefont{Rice}},
  \bibinfo{journal}{Annals of Physics} \textbf{\bibinfo{volume}{324}},
  \bibinfo{pages}{1452} (\bibinfo{year}{2009}).

\bibitem[{\citenamefont{Wang et~al.}(2013)\citenamefont{Wang, Li, Xiang, and
  Wang}}]{wang:2013}
\bibinfo{author}{\bibfnamefont{W.-S.} \bibnamefont{Wang}},
  \bibinfo{author}{\bibfnamefont{Z.-Z.} \bibnamefont{Li}},
  \bibinfo{author}{\bibfnamefont{Y.-Y.} \bibnamefont{Xiang}}, \bibnamefont{and}
  \bibinfo{author}{\bibfnamefont{Q.-H.} \bibnamefont{Wang}},
  \bibinfo{journal}{Physical Review B} \textbf{\bibinfo{volume}{87}},
  \bibinfo{pages}{115135} (\bibinfo{year}{2013}).

\bibitem[{\citenamefont{Gonzalez}(2008)}]{gonzalez:2008}
\bibinfo{author}{\bibfnamefont{J.}~\bibnamefont{Gonzalez}},
  \bibinfo{journal}{Physical Review B} \textbf{\bibinfo{volume}{78}},
  \bibinfo{pages}{205431} (\bibinfo{year}{2008}).

\bibitem[{\citenamefont{Martin and Batista}(2008)}]{martin:2008}
\bibinfo{author}{\bibfnamefont{I.}~\bibnamefont{Martin}} \bibnamefont{and}
  \bibinfo{author}{\bibfnamefont{C.}~\bibnamefont{Batista}},
  \bibinfo{journal}{Physical review letters} \textbf{\bibinfo{volume}{101}},
  \bibinfo{pages}{156402} (\bibinfo{year}{2008}).

\bibitem[{\citenamefont{Halboth and Metzner}(2000)}]{halboth:2000}
\bibinfo{author}{\bibfnamefont{C.~J.} \bibnamefont{Halboth}} \bibnamefont{and}
  \bibinfo{author}{\bibfnamefont{W.}~\bibnamefont{Metzner}},
  \bibinfo{journal}{Physical Review B} \textbf{\bibinfo{volume}{61}},
  \bibinfo{pages}{7364} (\bibinfo{year}{2000}).

\bibitem[{\citenamefont{Honerkamp and Salmhofer}(2001)}]{honerkamp:2001t}
\bibinfo{author}{\bibfnamefont{C.}~\bibnamefont{Honerkamp}} \bibnamefont{and}
  \bibinfo{author}{\bibfnamefont{M.}~\bibnamefont{Salmhofer}},
  \bibinfo{journal}{Physical Review B} \textbf{\bibinfo{volume}{64}},
  \bibinfo{pages}{184516} (\bibinfo{year}{2001}).

\bibitem[{\citenamefont{Schulz}(1987)}]{schulz:1987}
\bibinfo{author}{\bibfnamefont{H.}~\bibnamefont{Schulz}},
  \bibinfo{journal}{Europhysics Letters} \textbf{\bibinfo{volume}{4}},
  \bibinfo{pages}{609} (\bibinfo{year}{1987}).

\bibitem[{\citenamefont{Kiesel and Thomale}(2012)}]{kiesel:2012s}
\bibinfo{author}{\bibfnamefont{M.~L.} \bibnamefont{Kiesel}} \bibnamefont{and}
  \bibinfo{author}{\bibfnamefont{R.}~\bibnamefont{Thomale}},
  \bibinfo{journal}{Physical Review B} \textbf{\bibinfo{volume}{86}},
  \bibinfo{pages}{121105} (\bibinfo{year}{2012}).

\bibitem[{\citenamefont{Yu and Li}(2012)}]{yu:2012}
\bibinfo{author}{\bibfnamefont{S.-L.} \bibnamefont{Yu}} \bibnamefont{and}
  \bibinfo{author}{\bibfnamefont{J.-X.} \bibnamefont{Li}},
  \bibinfo{journal}{Physical Review B} \textbf{\bibinfo{volume}{85}},
  \bibinfo{pages}{144402} (\bibinfo{year}{2012}).

\bibitem[{\citenamefont{Classen et~al.}(2020)\citenamefont{Classen, Chubukov,
  Honerkamp, and Scherer}}]{classen:2020}
\bibinfo{author}{\bibfnamefont{L.}~\bibnamefont{Classen}},
  \bibinfo{author}{\bibfnamefont{A.~V.} \bibnamefont{Chubukov}},
  \bibinfo{author}{\bibfnamefont{C.}~\bibnamefont{Honerkamp}},
  \bibnamefont{and} \bibinfo{author}{\bibfnamefont{M.~M.}
  \bibnamefont{Scherer}}, \bibinfo{journal}{Physical Review B}
  \textbf{\bibinfo{volume}{102}}, \bibinfo{pages}{125141}
  (\bibinfo{year}{2020}).

\bibitem[{\citenamefont{McKenzie}(1997)}]{mckenzie:1997}
\bibinfo{author}{\bibfnamefont{R.~H.} \bibnamefont{McKenzie}},
  \bibinfo{journal}{Science} \textbf{\bibinfo{volume}{278}},
  \bibinfo{pages}{820} (\bibinfo{year}{1997}).

\bibitem[{\citenamefont{Williams et~al.}(1990)\citenamefont{Williams, Kini,
  Wang, Carlson, Geiser, Montgomery, Pyrka, Watkins, and
  Kommers}}]{williams:1990}
\bibinfo{author}{\bibfnamefont{J.~M.} \bibnamefont{Williams}},
  \bibinfo{author}{\bibfnamefont{A.~M.} \bibnamefont{Kini}},
  \bibinfo{author}{\bibfnamefont{H.~H.} \bibnamefont{Wang}},
  \bibinfo{author}{\bibfnamefont{K.~D.} \bibnamefont{Carlson}},
  \bibinfo{author}{\bibfnamefont{U.}~\bibnamefont{Geiser}},
  \bibinfo{author}{\bibfnamefont{L.~K.} \bibnamefont{Montgomery}},
  \bibinfo{author}{\bibfnamefont{G.~J.} \bibnamefont{Pyrka}},
  \bibinfo{author}{\bibfnamefont{D.~M.} \bibnamefont{Watkins}},
  \bibnamefont{and} \bibinfo{author}{\bibfnamefont{J.~M.}
  \bibnamefont{Kommers}}, \bibinfo{journal}{Inorganic Chemistry}
  \textbf{\bibinfo{volume}{29}}, \bibinfo{pages}{3272} (\bibinfo{year}{1990}).

\bibitem[{\citenamefont{Miyagawa et~al.}(2004)\citenamefont{Miyagawa, Kanoda,
  and Kawamoto}}]{miyagawa:2004}
\bibinfo{author}{\bibfnamefont{K.}~\bibnamefont{Miyagawa}},
  \bibinfo{author}{\bibfnamefont{K.}~\bibnamefont{Kanoda}}, \bibnamefont{and}
  \bibinfo{author}{\bibfnamefont{A.}~\bibnamefont{Kawamoto}},
  \bibinfo{journal}{Chemical reviews} \textbf{\bibinfo{volume}{104}},
  \bibinfo{pages}{5635} (\bibinfo{year}{2004}).

\bibitem[{\citenamefont{Kagawa et~al.}(2005)\citenamefont{Kagawa, Miyagawa, and
  Kanoda}}]{kagawa:2005}
\bibinfo{author}{\bibfnamefont{F.}~\bibnamefont{Kagawa}},
  \bibinfo{author}{\bibfnamefont{K.}~\bibnamefont{Miyagawa}}, \bibnamefont{and}
  \bibinfo{author}{\bibfnamefont{K.}~\bibnamefont{Kanoda}},
  \bibinfo{journal}{Nature} \textbf{\bibinfo{volume}{436}},
  \bibinfo{pages}{534} (\bibinfo{year}{2005}).

\bibitem[{\citenamefont{Kurosaki et~al.}(2005)\citenamefont{Kurosaki, Shimizu,
  Miyagawa, Kanoda, and Saito}}]{kurosaki:2005}
\bibinfo{author}{\bibfnamefont{Y.}~\bibnamefont{Kurosaki}},
  \bibinfo{author}{\bibfnamefont{Y.}~\bibnamefont{Shimizu}},
  \bibinfo{author}{\bibfnamefont{K.}~\bibnamefont{Miyagawa}},
  \bibinfo{author}{\bibfnamefont{K.}~\bibnamefont{Kanoda}}, \bibnamefont{and}
  \bibinfo{author}{\bibfnamefont{G.}~\bibnamefont{Saito}},
  \bibinfo{journal}{Physical review letters} \textbf{\bibinfo{volume}{95}},
  \bibinfo{pages}{177001} (\bibinfo{year}{2005}).

\bibitem[{\citenamefont{Miyagawa et~al.}(2002)\citenamefont{Miyagawa, Kawamoto,
  and Kanoda}}]{miyagawa:2002}
\bibinfo{author}{\bibfnamefont{K.}~\bibnamefont{Miyagawa}},
  \bibinfo{author}{\bibfnamefont{A.}~\bibnamefont{Kawamoto}}, \bibnamefont{and}
  \bibinfo{author}{\bibfnamefont{K.}~\bibnamefont{Kanoda}},
  \bibinfo{journal}{Physical review letters} \textbf{\bibinfo{volume}{89}},
  \bibinfo{pages}{017003} (\bibinfo{year}{2002}).

\bibitem[{\citenamefont{Kanoda}(1997)}]{kanoda:1997}
\bibinfo{author}{\bibfnamefont{K.}~\bibnamefont{Kanoda}},
  \bibinfo{journal}{Physica C: Superconductivity}
  \textbf{\bibinfo{volume}{282}}, \bibinfo{pages}{299} (\bibinfo{year}{1997}).

\bibitem[{\citenamefont{Mayaffre et~al.}(1995)\citenamefont{Mayaffre, Wzietek,
  J{\'e}rome, Lenoir, and Batail}}]{mayaffre:1995}
\bibinfo{author}{\bibfnamefont{H.}~\bibnamefont{Mayaffre}},
  \bibinfo{author}{\bibfnamefont{P.}~\bibnamefont{Wzietek}},
  \bibinfo{author}{\bibfnamefont{D.}~\bibnamefont{J{\'e}rome}},
  \bibinfo{author}{\bibfnamefont{C.}~\bibnamefont{Lenoir}}, \bibnamefont{and}
  \bibinfo{author}{\bibfnamefont{P.}~\bibnamefont{Batail}},
  \bibinfo{journal}{Physical review letters} \textbf{\bibinfo{volume}{75}},
  \bibinfo{pages}{4122} (\bibinfo{year}{1995}).

\bibitem[{\citenamefont{De~Soto et~al.}(1995)\citenamefont{De~Soto, Slichter,
  Kini, Wang, Geiser, and Williams}}]{de:1995}
\bibinfo{author}{\bibfnamefont{S.~M.} \bibnamefont{De~Soto}},
  \bibinfo{author}{\bibfnamefont{C.~P.} \bibnamefont{Slichter}},
  \bibinfo{author}{\bibfnamefont{A.~M.} \bibnamefont{Kini}},
  \bibinfo{author}{\bibfnamefont{H.}~\bibnamefont{Wang}},
  \bibinfo{author}{\bibfnamefont{U.}~\bibnamefont{Geiser}}, \bibnamefont{and}
  \bibinfo{author}{\bibfnamefont{J.}~\bibnamefont{Williams}},
  \bibinfo{journal}{Physical Review B} \textbf{\bibinfo{volume}{52}},
  \bibinfo{pages}{10364} (\bibinfo{year}{1995}).

\bibitem[{\citenamefont{Kanoda et~al.}(1996)\citenamefont{Kanoda, Miyagawa,
  Kawamoto, and Nakazawa}}]{kanoda:1996}
\bibinfo{author}{\bibfnamefont{K.}~\bibnamefont{Kanoda}},
  \bibinfo{author}{\bibfnamefont{K.}~\bibnamefont{Miyagawa}},
  \bibinfo{author}{\bibfnamefont{A.}~\bibnamefont{Kawamoto}}, \bibnamefont{and}
  \bibinfo{author}{\bibfnamefont{Y.}~\bibnamefont{Nakazawa}},
  \bibinfo{journal}{Physical Review B} \textbf{\bibinfo{volume}{54}},
  \bibinfo{pages}{76} (\bibinfo{year}{1996}).

\bibitem[{\citenamefont{Urayama et~al.}(1988)\citenamefont{Urayama, Yamochi,
  Saito, Nozawa, Sugano, Kinoshita, Sato, Oshima, Kawamoto, and
  Tanaka}}]{urayama:1988}
\bibinfo{author}{\bibfnamefont{H.}~\bibnamefont{Urayama}},
  \bibinfo{author}{\bibfnamefont{H.}~\bibnamefont{Yamochi}},
  \bibinfo{author}{\bibfnamefont{G.}~\bibnamefont{Saito}},
  \bibinfo{author}{\bibfnamefont{K.}~\bibnamefont{Nozawa}},
  \bibinfo{author}{\bibfnamefont{T.}~\bibnamefont{Sugano}},
  \bibinfo{author}{\bibfnamefont{M.}~\bibnamefont{Kinoshita}},
  \bibinfo{author}{\bibfnamefont{S.}~\bibnamefont{Sato}},
  \bibinfo{author}{\bibfnamefont{K.}~\bibnamefont{Oshima}},
  \bibinfo{author}{\bibfnamefont{A.}~\bibnamefont{Kawamoto}}, \bibnamefont{and}
  \bibinfo{author}{\bibfnamefont{J.}~\bibnamefont{Tanaka}},
  \bibinfo{journal}{Chemistry Letters} \textbf{\bibinfo{volume}{17}},
  \bibinfo{pages}{55} (\bibinfo{year}{1988}).

\bibitem[{\citenamefont{Kini et~al.}(1990)\citenamefont{Kini, Geiser, Wang,
  Carlson, Williams, Kwok, Vandervoort, Thompson, and Stupka}}]{kini:1990}
\bibinfo{author}{\bibfnamefont{A.~M.} \bibnamefont{Kini}},
  \bibinfo{author}{\bibfnamefont{U.}~\bibnamefont{Geiser}},
  \bibinfo{author}{\bibfnamefont{H.~H.} \bibnamefont{Wang}},
  \bibinfo{author}{\bibfnamefont{K.~D.} \bibnamefont{Carlson}},
  \bibinfo{author}{\bibfnamefont{J.~M.} \bibnamefont{Williams}},
  \bibinfo{author}{\bibfnamefont{W.}~\bibnamefont{Kwok}},
  \bibinfo{author}{\bibfnamefont{K.}~\bibnamefont{Vandervoort}},
  \bibinfo{author}{\bibfnamefont{J.~E.} \bibnamefont{Thompson}},
  \bibnamefont{and} \bibinfo{author}{\bibfnamefont{D.~L.}
  \bibnamefont{Stupka}}, \bibinfo{journal}{Inorganic Chemistry}
  \textbf{\bibinfo{volume}{29}}, \bibinfo{pages}{2555} (\bibinfo{year}{1990}).

\bibitem[{\citenamefont{Kuroki et~al.}(2002)\citenamefont{Kuroki, Kimura,
  Arita, Tanaka, and Matsuda}}]{kuroki:2002}
\bibinfo{author}{\bibfnamefont{K.}~\bibnamefont{Kuroki}},
  \bibinfo{author}{\bibfnamefont{T.}~\bibnamefont{Kimura}},
  \bibinfo{author}{\bibfnamefont{R.}~\bibnamefont{Arita}},
  \bibinfo{author}{\bibfnamefont{Y.}~\bibnamefont{Tanaka}}, \bibnamefont{and}
  \bibinfo{author}{\bibfnamefont{Y.}~\bibnamefont{Matsuda}},
  \bibinfo{journal}{Physical Review B} \textbf{\bibinfo{volume}{65}},
  \bibinfo{pages}{100516} (\bibinfo{year}{2002}).

\bibitem[{\citenamefont{Sekine et~al.}(2013)\citenamefont{Sekine, Nasu, and
  Ishihara}}]{sekine:2013}
\bibinfo{author}{\bibfnamefont{A.}~\bibnamefont{Sekine}},
  \bibinfo{author}{\bibfnamefont{J.}~\bibnamefont{Nasu}}, \bibnamefont{and}
  \bibinfo{author}{\bibfnamefont{S.}~\bibnamefont{Ishihara}},
  \bibinfo{journal}{Physical Review B—Condensed Matter and Materials Physics}
  \textbf{\bibinfo{volume}{87}}, \bibinfo{pages}{085133}
  (\bibinfo{year}{2013}).

\bibitem[{\citenamefont{Guterding et~al.}(2016)\citenamefont{Guterding,
  Altmeyer, Jeschke, and Valent{\'\i}}}]{guterding:2016}
\bibinfo{author}{\bibfnamefont{D.}~\bibnamefont{Guterding}},
  \bibinfo{author}{\bibfnamefont{M.}~\bibnamefont{Altmeyer}},
  \bibinfo{author}{\bibfnamefont{H.~O.} \bibnamefont{Jeschke}},
  \bibnamefont{and}
  \bibinfo{author}{\bibfnamefont{R.}~\bibnamefont{Valent{\'\i}}},
  \bibinfo{journal}{Physical Review B} \textbf{\bibinfo{volume}{94}},
  \bibinfo{pages}{024515} (\bibinfo{year}{2016}).

\bibitem[{\citenamefont{Koretsune and Hotta}(2014)}]{koretsune:2014}
\bibinfo{author}{\bibfnamefont{T.}~\bibnamefont{Koretsune}} \bibnamefont{and}
  \bibinfo{author}{\bibfnamefont{C.}~\bibnamefont{Hotta}},
  \bibinfo{journal}{Physical Review B} \textbf{\bibinfo{volume}{89}},
  \bibinfo{pages}{045102} (\bibinfo{year}{2014}).

\bibitem[{\citenamefont{Kawasugi et~al.}(2016)\citenamefont{Kawasugi, Seki,
  Edagawa, Sato, Pu, Takenobu, Yunoki, Yamamoto, and Kato}}]{kawasugi:2016}
\bibinfo{author}{\bibfnamefont{Y.}~\bibnamefont{Kawasugi}},
  \bibinfo{author}{\bibfnamefont{K.}~\bibnamefont{Seki}},
  \bibinfo{author}{\bibfnamefont{Y.}~\bibnamefont{Edagawa}},
  \bibinfo{author}{\bibfnamefont{Y.}~\bibnamefont{Sato}},
  \bibinfo{author}{\bibfnamefont{J.}~\bibnamefont{Pu}},
  \bibinfo{author}{\bibfnamefont{T.}~\bibnamefont{Takenobu}},
  \bibinfo{author}{\bibfnamefont{S.}~\bibnamefont{Yunoki}},
  \bibinfo{author}{\bibfnamefont{H.~M.} \bibnamefont{Yamamoto}},
  \bibnamefont{and} \bibinfo{author}{\bibfnamefont{R.}~\bibnamefont{Kato}},
  \bibinfo{journal}{Nature communications} \textbf{\bibinfo{volume}{7}},
  \bibinfo{pages}{12356} (\bibinfo{year}{2016}).

\bibitem[{\citenamefont{Suh et~al.}(2023)\citenamefont{Suh, Yu, Shishidou,
  Weinert, Brydon, and Agterberg}}]{suh:2023}
\bibinfo{author}{\bibfnamefont{H.~G.} \bibnamefont{Suh}},
  \bibinfo{author}{\bibfnamefont{Y.}~\bibnamefont{Yu}},
  \bibinfo{author}{\bibfnamefont{T.}~\bibnamefont{Shishidou}},
  \bibinfo{author}{\bibfnamefont{M.}~\bibnamefont{Weinert}},
  \bibinfo{author}{\bibfnamefont{P.}~\bibnamefont{Brydon}}, \bibnamefont{and}
  \bibinfo{author}{\bibfnamefont{D.~F.} \bibnamefont{Agterberg}},
  \bibinfo{journal}{Physical Review Research} \textbf{\bibinfo{volume}{5}},
  \bibinfo{pages}{033204} (\bibinfo{year}{2023}).

\bibitem[{\citenamefont{Arovas et~al.}(2022)\citenamefont{Arovas, Berg,
  Kivelson, and Raghu}}]{arovas:2022}
\bibinfo{author}{\bibfnamefont{D.~P.} \bibnamefont{Arovas}},
  \bibinfo{author}{\bibfnamefont{E.}~\bibnamefont{Berg}},
  \bibinfo{author}{\bibfnamefont{S.~A.} \bibnamefont{Kivelson}},
  \bibnamefont{and} \bibinfo{author}{\bibfnamefont{S.}~\bibnamefont{Raghu}},
  \bibinfo{journal}{Annual review of condensed matter physics}
  \textbf{\bibinfo{volume}{13}}, \bibinfo{pages}{239} (\bibinfo{year}{2022}).

\bibitem[{\citenamefont{Krempask{\`y} et~al.}(2024)\citenamefont{Krempask{\`y},
  {\v{S}}mejkal, D’souza, Hajlaoui, Springholz, Uhl{\'\i}{\v{r}}ov{\'a},
  Alarab, Constantinou, Strocov, Usanov et~al.}}]{krempasky:2024}
\bibinfo{author}{\bibfnamefont{J.}~\bibnamefont{Krempask{\`y}}},
  \bibinfo{author}{\bibfnamefont{L.}~\bibnamefont{{\v{S}}mejkal}},
  \bibinfo{author}{\bibfnamefont{S.}~\bibnamefont{D’souza}},
  \bibinfo{author}{\bibfnamefont{M.}~\bibnamefont{Hajlaoui}},
  \bibinfo{author}{\bibfnamefont{G.}~\bibnamefont{Springholz}},
  \bibinfo{author}{\bibfnamefont{K.}~\bibnamefont{Uhl{\'\i}{\v{r}}ov{\'a}}},
  \bibinfo{author}{\bibfnamefont{F.}~\bibnamefont{Alarab}},
  \bibinfo{author}{\bibfnamefont{P.}~\bibnamefont{Constantinou}},
  \bibinfo{author}{\bibfnamefont{V.}~\bibnamefont{Strocov}},
  \bibinfo{author}{\bibfnamefont{D.}~\bibnamefont{Usanov}},
  \bibnamefont{et~al.}, \bibinfo{journal}{Nature}
  \textbf{\bibinfo{volume}{626}}, \bibinfo{pages}{517} (\bibinfo{year}{2024}).

\bibitem[{\citenamefont{Mazin et~al.}(2023)\citenamefont{Mazin,
  Gonz{\'a}lez-Hern{\'a}ndez, and {\v{S}}mejkal}}]{mazin:2023}
\bibinfo{author}{\bibfnamefont{I.}~\bibnamefont{Mazin}},
  \bibinfo{author}{\bibfnamefont{R.}~\bibnamefont{Gonz{\'a}lez-Hern{\'a}ndez}},
  \bibnamefont{and}
  \bibinfo{author}{\bibfnamefont{L.}~\bibnamefont{{\v{S}}mejkal}},
  \bibinfo{journal}{arXiv preprint arXiv:2309.02355}  (\bibinfo{year}{2023}).

\bibitem[{\citenamefont{Miyagawa et~al.}(1995)\citenamefont{Miyagawa, Kawamoto,
  Nakazawa, and Kanoda}}]{miyagawa:1995}
\bibinfo{author}{\bibfnamefont{K.}~\bibnamefont{Miyagawa}},
  \bibinfo{author}{\bibfnamefont{A.}~\bibnamefont{Kawamoto}},
  \bibinfo{author}{\bibfnamefont{Y.}~\bibnamefont{Nakazawa}}, \bibnamefont{and}
  \bibinfo{author}{\bibfnamefont{K.}~\bibnamefont{Kanoda}},
  \bibinfo{journal}{Physical review letters} \textbf{\bibinfo{volume}{75}},
  \bibinfo{pages}{1174} (\bibinfo{year}{1995}).

\bibitem[{\citenamefont{Milivojevi{\'c}
  et~al.}(2024)\citenamefont{Milivojevi{\'c}, Orozovi{\'c}, Picozzi, Gmitra,
  and Stavri{\'c}}}]{milivojevic:2024}
\bibinfo{author}{\bibfnamefont{M.}~\bibnamefont{Milivojevi{\'c}}},
  \bibinfo{author}{\bibfnamefont{M.}~\bibnamefont{Orozovi{\'c}}},
  \bibinfo{author}{\bibfnamefont{S.}~\bibnamefont{Picozzi}},
  \bibinfo{author}{\bibfnamefont{M.}~\bibnamefont{Gmitra}}, \bibnamefont{and}
  \bibinfo{author}{\bibfnamefont{S.}~\bibnamefont{Stavri{\'c}}},
  \bibinfo{journal}{2D Materials} \textbf{\bibinfo{volume}{11}},
  \bibinfo{pages}{035025} (\bibinfo{year}{2024}).

\bibitem[{\citenamefont{Guo et~al.}(2023{\natexlab{b}})\citenamefont{Guo, Liu,
  Janson, Fulga, van~den Brink, and Facio}}]{guo:2023S}
\bibinfo{author}{\bibfnamefont{Y.}~\bibnamefont{Guo}},
  \bibinfo{author}{\bibfnamefont{H.}~\bibnamefont{Liu}},
  \bibinfo{author}{\bibfnamefont{O.}~\bibnamefont{Janson}},
  \bibinfo{author}{\bibfnamefont{I.~C.} \bibnamefont{Fulga}},
  \bibinfo{author}{\bibfnamefont{J.}~\bibnamefont{van~den Brink}},
  \bibnamefont{and} \bibinfo{author}{\bibfnamefont{J.~I.} \bibnamefont{Facio}},
  \bibinfo{journal}{Materials Today Physics} \textbf{\bibinfo{volume}{32}},
  \bibinfo{pages}{100991} (\bibinfo{year}{2023}{\natexlab{b}}).

\bibitem[{\citenamefont{de~la Flor et~al.}(2021)\citenamefont{de~la Flor,
  Souvignier, Madariaga, and Aroyo}}]{de:2021}
\bibinfo{author}{\bibfnamefont{G.}~\bibnamefont{de~la Flor}},
  \bibinfo{author}{\bibfnamefont{B.}~\bibnamefont{Souvignier}},
  \bibinfo{author}{\bibfnamefont{G.}~\bibnamefont{Madariaga}},
  \bibnamefont{and} \bibinfo{author}{\bibfnamefont{M.~I.} \bibnamefont{Aroyo}},
  \bibinfo{journal}{Acta Crystallographica Section A: Foundations and Advances}
  \textbf{\bibinfo{volume}{77}}, \bibinfo{pages}{559} (\bibinfo{year}{2021}).

\bibitem[{\citenamefont{Kopsk{\`y} and Litvin}(2002)}]{kopsky:2002}
\bibinfo{author}{\bibfnamefont{V.}~\bibnamefont{Kopsk{\`y}}} \bibnamefont{and}
  \bibinfo{author}{\bibfnamefont{D.}~\bibnamefont{Litvin}}
  (\bibinfo{year}{2002}).

\end{thebibliography}

{\it Acknowledgements: }We thank Philip Brydon, Andrey Chubukov, Tatsuya Shishidou, Stephen Wilson, and Michael Weinert for useful discussions. D.F.A. and Y.Y were supported by the
National Science Foundation Grant No. DMREF 2323857 (for one loop and patch RG calculations). D.F.A. and H.G.S. were supported by the Department of Energy, Office of Basic Energy Science, Division of Materials Sciences and Engineering under Award
No. DE-SC0021971 (for the development of the symmetry-based Hamiltonians). M.~R. acknowledges support from the Novo Nordisk Foundation grant NNF20OC0060019.

{\it Author Contributions Statement:} Y.Y., H.G.S., M.R., and D.F.A. designed research; performed research; contributed analytic tools; and wrote the
paper.

{\it Competing Interests Statement: }The authors declare no competing interest.

\begin{table}[h]
\begin{tabular}{|c|c|}
\hline
2D& $\begin{matrix}
\text{L17(p2$_1$/b11), L21(p$2_12_1$2), L25(pba2)}\\ \text{L44(pbam), L54(p4$2_1$2), L56(p4bm)} \\
\text{L58(p$\overline{4}2_1$m), L60(p$\overline{4}$b2), L63(p4/mbm)}
\end{matrix}$ \\ \hline
3D&$\begin{matrix}
\text{18(P$2_12_1$2)R\&S, 19(P$2_12_12_1$)S\&T($k_yk_z$)\&U($k_xk_z$)}
\\ \text{55(Pbam)R\&S, 56(Pccn)R, 58(Pnnm)S}\\ \text{62(Pnma)U, 90(P42$_1$2)M\&A, 92(P4$_1$2$_1$2)M}\\ \text{94(P4$_2$2$_1$2)M\&A, 96(P4$_3$2$_1$2)M, 127(P4/mbm)M\&A}\\ \text{128(P4/mnc)M, 135(P4$_2$/mbc)M, 136(P4$_2$/mnm)M}\\ \text{138(P4$_2$/ncm)A, 212(P4$_3$32)M, 213(P4$_1$32)M}\end{matrix}$\\ \hline
\end{tabular}
\caption{{\red 2D layer groups and 3D space groups \cite{de:2021,kopsky:2002} hosting the kp Hamiltonian in Eq.\ref{Eq:kp} at quadratic level. For 2D layer groups, the kp Hamiltonian is centered at the $(\pi,\pi)$ point. For 3D space groups, the corresponding high-symmetry points are included. In 19T(19U), the coefficients of $\tau_{x,z}$ terms are $k_yk_z$($k_xk_z$) instead of $k_xk_y$.}}
\label{T:SG}
\end{table}

\clearpage
\appendix
\setcounter{figure}{0}
\renewcommand{\figurename}{Supplementary Fig.}

\section{Evaluating the self-consistent one-loop vertex corrections}\label{S:AC}
{\red In this section, we will analyze the instabilities from negative $g_1$ and $g_2$ through the self-consistent one-loop vertex correction.}

\begin{figure}[ht]
\centering
\includegraphics[height=2.5cm]{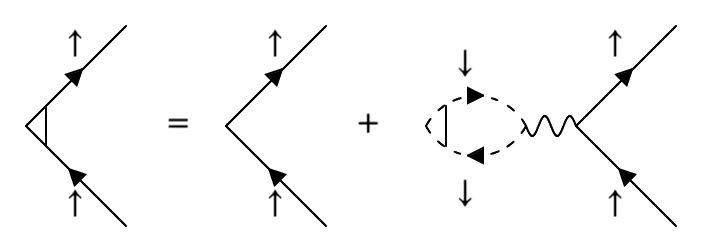}
\includegraphics[height=2.5cm]{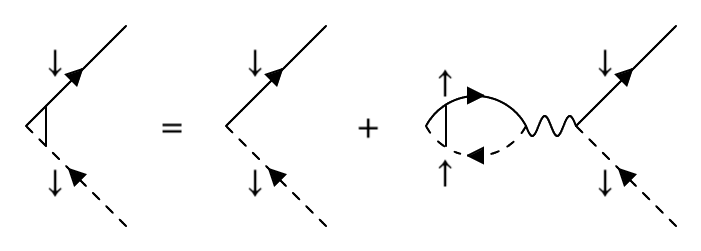}
\includegraphics[height=2.5cm]{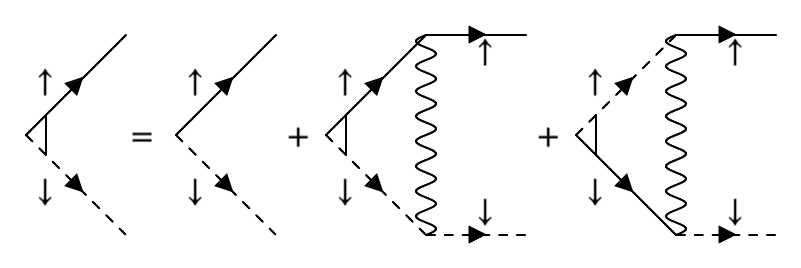}
\caption{Self-consistent one-loop vertex corrections. (Top) Intra-band particle-hole vertex correction for altermagnetism. (Middle) Inter-band particle-hole vertex correction for nematicity and orbital altermagnetism. (Bottom) Inter-band particle-particle vertex correction for inter-band superconductivity.}
\label{F:1loop3}
\end{figure}

For particle-hole vertices, we only consider the $(\uparrow,\uparrow)$ and $(\downarrow,\downarrow)$ vertices due to the spin-rotational symmetry. The top panel of Fig.\ref{F:1loop3} shows a self-consistent one-loop correction in the intra-band particle-hole channel. It describes the correction to the $\Gamma_{1\uparrow,1\uparrow}$ vertex from the $\Gamma_{2\downarrow,2\downarrow}$ vertex. {\red This diagram involves the $g_2$ interaction and the intra-band particle-hole susceptibility of the second band (dashed loop).} The self-consistent equation can be written as a matrix form:
\begin{equation}
\begin{split}
\vec{\Gamma}_{ph}^{intra}&\equiv\left[\Gamma_{1\uparrow,1\uparrow},\Gamma_{1\downarrow,1\downarrow},\Gamma_{2\uparrow,2\uparrow},\Gamma_{2\downarrow,2\downarrow}\right]\\
\vec{\Gamma}_{ph}^{intra}&=\vec{\Gamma}_{ph,0}^{intra}+M_{ph}^{intra}\vec{\Gamma}_{ph}^{intra} \\
M_{ph}^{intra}&=-\chi_{ph}^{intra}g_2\left[\begin{array}{cccc}
     &&& 1 \\
     &&1& \\
     &1&& \\
     1&&& 
\end{array}\right]
\end{split}
\end{equation}
{\red The shown diagram describes the first row of the matrix.} The leading instability in this channel corresponds to the eigenvector of the matrix $M_{ph}^{intra}$ with the largest positive eigenvalue. For a negative $g_2$, such eigenvector is $[1,-1,-1,1]$ with eigenvalue $\chi_{ph}^{intra}|g_2|$. The corresponding vertex, $\Gamma_{1\uparrow,1\uparrow}-\Gamma_{1\downarrow,1\downarrow}-\Gamma_{2\uparrow,2\uparrow}+\Gamma_{2\downarrow,2\downarrow}$, describes the altermagnetic state $M=\widetilde{\tau_z}\sigma_z$. Another eigenvector $[1,1,1,1]$ has the same eigenvalue but does not correspond to any orderings. 

The middle panel of Fig.\ref{F:1loop3} shows a self-consistent one-loop correction in the inter-band particle-hole channel. It describes the correction to the $\Gamma_{1\downarrow,2\downarrow}$ vertex from the $\Gamma_{1\uparrow,2\uparrow}$ vertex. {\red This diagram involves the $g_1$ interaction and the inter-band particle-hole susceptibility (solid-dashed loop).} The self-consistent equations of this channel have the following matrix form:
\begin{equation}
\begin{split}
\vec{\Gamma}_{ph}^{inter}&\equiv\left[\Gamma_{1\uparrow,2\uparrow},\Gamma_{1\downarrow,2\downarrow},\Gamma_{2\uparrow,1\uparrow},\Gamma_{2\downarrow,1\downarrow}\right]\\
\vec{\Gamma}_{ph}^{inter}&=\vec{\Gamma}_{ph,0}^{inter}+M_{ph}^{inter}\vec{\Gamma}_{ph}^{inter} \\
M_{ph}^{inter}&=-\chi_{ph}^{inter}g_1\left[\begin{array}{cccc}
     &1&& \\
     1&&& \\
     &&&1 \\
     &&1& 
\end{array}\right]
\end{split}
\end{equation}
{\red The shown diagram describes the second row of the matrix.} There are two leading instabilities for $g_1<0$ with the same eigenvalue $\chi_{ph}^{inter}|g_1|$. The first one has eigenvector $[1,1,1,1]$. The corresponding vertex, $\Gamma_{1\uparrow,2\uparrow}+\Gamma_{1\downarrow,2\downarrow}+\Gamma_{2\uparrow,1\uparrow}+\Gamma_{2\downarrow,1\downarrow}$, describes the nematicity $O=\widetilde{\tau_x}$. In the lattice basis, it has components $\tau_x$ and $\tau_z$ with the same symmetry as $k_xk_y$, so it breaks mirror reflections symmetries for $\widetilde{M_x}$ and $\widetilde{M_y}$. The second instability has eigenvector $[1,1,-1,-1]$. The corresponding vertex, $\Gamma_{1\uparrow,2\uparrow}+\Gamma_{1\downarrow,2\downarrow}-\Gamma_{2\uparrow,1\uparrow}-\Gamma_{2\downarrow,1\downarrow}$ describes another unconventional magnetic state $O=\widetilde{\tau_y}$. In the lattice basis, it is $\tau_y$, which is an even-parity current-loop order. {\red This current loop order only breaks the time-reversal symmetry while preserving all crystal reflection symmetries.} When SOC is present, it has the same symmetry as the altermagnetism $k_xk_y\sigma_z$.

The bottom panel of Fig.\ref{F:1loop3} shows a self-consistent one-loop correction in the inter-band particle-particle channel. {\red It captures two corrections to the $\Gamma^{sc}_{1\uparrow,2\downarrow}$ vertex. The first correction is from $\Gamma^{sc}_{1\uparrow,2\downarrow}$ and involves the $g_2$ interaction. The second correction is from $\Gamma^{sc}_{2\uparrow,1\downarrow}$ and involves the $g_1$ interaction. Both corrections involve the inter-band particle-particle susceptibility $\chi_{pp}^{inter}$.} The self-consistent equations in this channel have a matrix form:
\begin{equation}
\begin{split}
\vec{\Gamma}_{pp}^{inter}&\equiv\left[\Gamma^{sc}_{1\uparrow,2\downarrow},\Gamma^{sc}_{2\uparrow,1\downarrow}\right]\\
\vec{\Gamma}_{pp}^{inter}&=\vec{\Gamma}_{pp,0}^{inter}+M_{pp}^{inter}\vec{\Gamma}_{pp}^{inter} \\
M_{pp}^{inter}&=-\chi_{pp}^{inter}\left[\begin{array}{cc}
     g_2&g_1 \\
     g_1&g_2 
\end{array}\right]
\end{split}
\end{equation}
{\red The shown diagram describes the first row of the matrix.} For $g_1,g_2<0$, the eigenvector with the largest positive eigenvalue is $[1,1]$, with eigenvalue $\chi_{pp}^{inter}|g_1+g_2|$. The corresponding vertex, $\Gamma^{sc}_{1\uparrow,2\downarrow}+\Gamma^{sc}_{2\uparrow,1\downarrow}$, describes the inter-band superconductivity $\Delta=\widetilde{\tau_x}i\sigma_y$. In the lattice basis, it has components $\tau_xi\sigma_y$ and $\tau_zi\sigma_y$, with the same symmetry as a d-wave superconductor $k_xk_yi\sigma_y$. There is no vertex correction in the intra-band particle-particle channel from $g_1$ or $g_2$.

\section{Patch RG Calculation}
We perform a one-loop patch renormalization group analysis. When only keeping the leading (BCS) correction, the RG flow is
\begin{equation}
\begin{split}
&\dot{g_1}=-2\dot{\chi}_{pp}^{intra}g_5^2\\
&\dot{g_2}=-2\dot{\chi}_{pp}^{intra}g_5^2\\
&\dot{g_4}+\dot{g_3}=-\dot{\chi}_{pp}^{intra}(g_4+g_3)^2\\
&\dot{g_4}-\dot{g_3}=-\dot{\chi}_{pp}^{intra}(g_4-g_3)^2\\
&\dot{g_5}=\dot{\chi}_{pp}^{intra}g_5(g_3-g_4)
\end{split}
\end{equation}
Here $\dot{...}\equiv \frac{d(...)}{dt}$, with $t\equiv\log\Lambda/T$. Note that $g_5$ does not introduce BCS correction to $g_{3,4}$. The RG flow for $g_{3,4}$ has three fixed points. The first fixed point has divergent $g_4+g_3<0$ while finite $g_4-g_3$, describing the standard BCS instability for s-wave superconductivity $\Delta=i\sigma_y$. This fixed point is not affected by $g_5$, as $g_5$ is finite. 
The second fixed point has divergent $g_4-g_3<0$ while finite $g_4+g_3$, originally describing the BCS instability for d-wave superconductivity $\Delta=i\tau_z\sigma_y$. This fixed point has divergent $g_5$, and consequently, divergent $g_{1,2}<0$. The presence of $g_5$ leads to other subleading instabilities.
In this work, we focus on the last fixed point scenario where $g_4$ starts sufficiently repulsive, such that $g_4\pm g_3$ are both reduced under the RG flow. 

The full RG flows are :
\begin{equation}
\begin{split}
\dot{g_1}&=2g_1g_4+2ag_1(g_2-g_1)-2bg_1g_2\\&-(2ct+2a)g_5^2\\
\dot{g_2}&=2(g_1-g_2)g_4+a(g_2^2+g_3^2)-b(g_1^2+g_2^2)\\&-(2ct-2)g_5^2\\
\dot{g_3}&=2a(2g_2-g_1)g_3-2ctg_3g_4+(2b+2)g_5^2\\
\dot{g_4}&=g_1^2-2g_2^2+2g_1g_2+g_4^2-ct(g_3^2+g_4^2)\\&+2(a-b)g_5^2\\
\dot{g_5}&=g_5\left[g_2-2g_1+a(2g_2-g_1)-b(g_1+g_2)\right.\\&\left.+g_4+ag_3+ct(g_3-g_4)\right]
\end{split}
\end{equation}
For simplicity, we have defined $\dot{\chi}_{ph}^{intra}\equiv1$, $\dot{\chi}_{ph}^{inter}\equiv a$, $\dot{\chi}_{pp}^{inter}\equiv b$, and $\dot{\chi}_{pp}^{intra}\equiv ct$. 

Let us focus on the sector of fixed point, where $g_{3,4}$ becomes small due to their BCS corrections. The RG flow can be simplified to:
\begin{equation}
\begin{split}
&\dot{g_1}=2ag_1(g_2-g_1)-2bg_1g_2-2ctg_5^2\\
&\dot{g_2}=ag_2^2-b(g_1^2+g_2^2)-2ctg_5^2\\
&\dot{g_5}=g_5\left[g_2-2g_1+a(2g_2-g_1)-b(g_1+g_2)\right]
\end{split}
\end{equation}
{\red We will come back to $g_{3,4}$ when analyzing the fixed point solution. For the $g_5^2$ corrections, we only keep the dominant intra-band particle-particle channel. Now the theory contains both contributions from $\chi_{ph}^{intra}$, $\chi_{ph}^{inter}$, and $\chi_{pp}^{inter}$; as well as the stronger $\chi_{pp}^{intra}$. They are the leading corrections from $g_{1,2}$ and $g_5$, respectively. Notably, these susceptibilities differ by a factor of $\log\Lambda/T$. }

Near instabilities $t\rightarrow t_c$, the RG flow for interaction strengths diverges. When studying weak-coupling instabilities, the change in the interaction strengths predominantly happens near $t\rightarrow t_c$. As an approximation, we will take $t=t_c$ in the RG flow. Now the BCS corrections all have the form $ct_c g_5^2$. {\red Due to the special form of the RG flow for $g_5$, }we can define $\widetilde{g_5}=\sqrt{ct_c}g_5$ to absorb the tuning paramter $t_c$. The RG flow becomes
\begin{equation}
\begin{split}
&\dot{g_1}=2ag_1(g_2-g_1)-2bg_1g_2-2\widetilde{g_5}^2\\
&\dot{g_2}=ag_2^2-b(g_1^2+g_2^2)-2\widetilde{g_5}^2\\
&\dot{\widetilde{g_5}}=\widetilde{g_5}\left[g_2-2g_1+a(2g_2-g_1)-b(g_1+g_2)\right]
\end{split}
\end{equation}
At the end of the calculation, we will take the $t_c\gg1$ limit. For the above equations, the generic solution for an interaction strength X is $X\rightarrow\frac{G_X}{t_c-t}$. {\red Near the fixed point, $g_1$, $g_2$, and $\widetilde{g_5}=\sqrt{ct_c}g_5$ are then on the same order.} The equations for $G_X$ are:
\begin{equation}
\begin{split}
&G_1=2aG_1(G_2-G_1)-2bG_1G_2-2\widetilde{G_5}^2\\
&G_2=aG_2^2-b(G_1^2+G_2^2)-2\widetilde{G_5}^2\\
&\widetilde{G_5}=\widetilde{G_5}\left[G_2-2G_1+a(2G_2-G_1)-b(G_1+G_2)\right]
\end{split}
\end{equation}
To understand the dominant instability at a given fixed point, we turn to the vertex RG. The RG flow for a given vertex $\Gamma$ reads $\dot{\Gamma}=M\Gamma$, where $M$ is a linear function of $X$. The generic solution for the vertices is $\Gamma\propto(t_c-t)^{-\gamma_i}$. The corresponding susceptiblity scales as $\chi_\Gamma\propto(t_c-t)^{1-2\gamma_i}$. The leading instability thus corresponds to the largest positive $\gamma_i$. These powers $\gamma_i$ are listed in Table. \ref{T:A1}. Here, we have taken $t_c\gg1$, such that $\widetilde{G_5}$ is kept, while $\widetilde{G_5}/\sqrt{t_c}$ is neglected. {\red 
Notably, even though $G_5\propto \widetilde{G_5}/\sqrt{t_c}\ll G_{1,2}$, its contribution on the SC instability cannot be neglected, because the corresponding vertex RG flow is big. We will explicitly show this effect in the superconducting channel below.

The RG flow for the SC vertices with opposite spin pairing $\Gamma_{SC}=\left[\Gamma_{1\uparrow,1\downarrow},\Gamma_{2\uparrow,2\downarrow},\Gamma_{1\uparrow,2\downarrow},\Gamma_{2\uparrow,1\downarrow}\right]$ is:
\begin{equation}
\begin{split}
\dot{\Gamma_{SC}}=\left[\begin{array}{cccc}
     0&0&-bg_5&-bg_5 \\
     0&0&bg_5&bg_5\\
     -ct_cg_5&ct_cg_5&-bg_2&-bg_1\\
     -ct_cg_5&ct_cg_5&-bg_1&-bg_2
\end{array}\right]\Gamma_{SC}
\end{split}
\label{E:vertexRG}
\end{equation}
}
\begin{figure}[h]
\centering
\includegraphics[height=2.2cm]{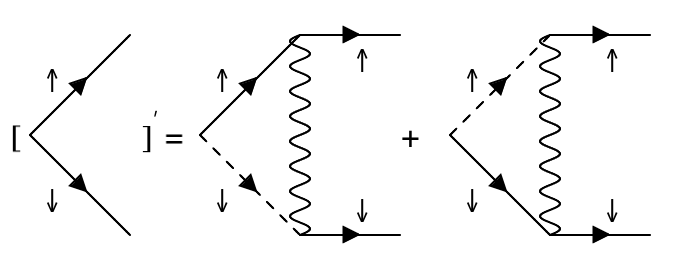}
\includegraphics[height=2.2cm]{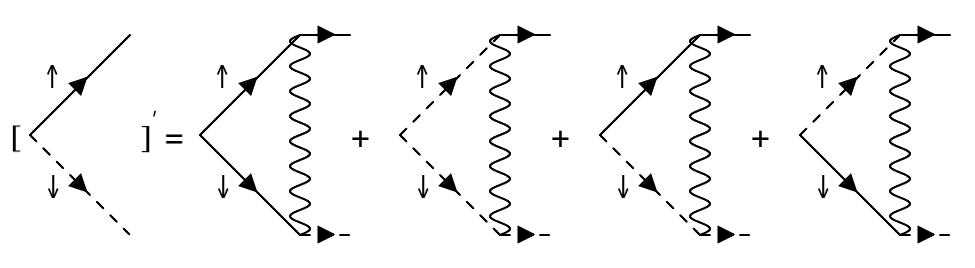}
\caption{Vertex corrections in the SC (particle-particle) channel from $g_{1,2,5}$.}
\label{F:vertexRG}
\end{figure}

{\red The corresponding Feynman diagrams are shown in Fig.\ref{F:vertexRG}. The top panel gives the first row of the matrix in Eq.\ref{E:vertexRG}.
It describes how $g_5$ interaction corrects the intra-band vertex $\Gamma_{1\uparrow,1\downarrow}$. Its contributions are from the inter-band vertices $\Gamma_{1\uparrow,2\downarrow}$ and $\Gamma_{2\uparrow,1\downarrow}$, through the the inter-band susceptibility $\dot{\chi}_{pp}^{inter}\equiv b$. The bottom panel gives the third row of the matrix in Eq.\ref{E:vertexRG}. It describes how $g_{1,2,5}$ corrects the inter-band $\Gamma_{1\uparrow,2\downarrow}$ vertex. The $g_5$ contributions are described by the first two diagrams. They are from the intra-band vertices $\Gamma_{1\uparrow,1\downarrow}$ and $\Gamma_{2\uparrow,2\downarrow}$, through the intra-band (BCS) susceptibility $\dot{\chi}_{pp}^{intra}\equiv ct_c$. The $g_{1,2}$ contributions are included as the last two diagrams, and they have been discussed in the main text and Sec.\ref{S:AC}. They describe corrections from the inter-band vertices $\Gamma_{1\uparrow,2\downarrow}$ and $\Gamma_{2\uparrow,1\downarrow}$, through the inter-band susceptibility. 

Similar to the interaction RG flow, the vertex RG flow also contains both inter-band and intra-band (BCS) susceptibilities, as they are the leading corrections from $g_{1,2}$ and $g_5$, respectively. The corresponding susceptibilities differ by a factor of $t_c=\log\Lambda/T_c$.

$\gamma$ depends on the eigenvalue of the matrix in Eq.\ref{E:vertexRG}. Simple s-wave state $\Gamma=[1,1,0,0]$ has zero eigenvalue, as $g_{3,4}$ are not considered. Orbit-singlet spin-triplet state $\Delta=\tau_y\sigma_z(i\sigma_y)$, with $\Gamma=[0,0,1,-1]$, has $\gamma=-b(G_2-G_1)$. Staggered singlet state $\Delta=\widetilde{\tau_z}(i\sigma_y)$ (with $\Gamma=[1,-1,0,0]$) and inter-band singlet state $\Delta=\widetilde{\tau_x}(i\sigma_y)$ (with $\Gamma=[0,0,1,1]$) mix up as they have the same symmetry. Their eigenvalues are $\frac{-b(G_2+G_1)}{2}\pm\sqrt{(\frac{- b(G_2+G_1)}{2})^2+4b\widetilde{G_5}^2}$. 
Notably, although $G_5 \ll G_{1,2}$ due to the interaction RG flow, their contributions to $\gamma$ are of the same order here because of the BCS correction of $g_5$ in the vertex RG flow. 

Formally, the tuning parameter $t_c$ does not appear in $\gamma$ when we took $t_c \to \infty$. However, the fixed point still captures both the diagrams with $\chi_{pp}^{intra}$ susceptibility, and the diagrams with $\chi_{ph}^{intra}$, $\chi_{ph}^{inter}$, and $\chi_{pp}^{inter}$ susceptibilities. The corresponding bare susceptibilities differ by a factor $\log\Lambda/T$. This is an unexpected result as such free tuning parameters usually strongly influence the final results, for example, in \cite{classen:2020}.}

\begin{table}[ht]
\begin{tabular}{|c|c|c|c|c|c|}
\hline
O   & $\sigma_z$ & $\widetilde{\tau_z}$      & $\widetilde{\tau_z}\sigma_z$ & $\widetilde{\tau_x}$          & $\widetilde{\tau_x}\sigma_z$  \\ \hline
$\gamma$ & $G_1$      & $2G_2-G_1$ & $-G_1$           & $a(-2G_1+G_2)$ & $aG_2$           \\ \hline
\end{tabular}

\begin{tabular}{|c|c|c|}
\hline
$\Delta$& $\widetilde{\tau_{x,z}}i\sigma_y$&$\tau_y\sigma_zi\sigma_y$\\ \hline
$\gamma$&$\frac{-b(G_2+G_1)}{2}\pm\sqrt{(\frac{- b(G_2+G_1)}{2})^2+4b\widetilde{G_5}^2}$&$-b(G_{2}-G_1)$\\
\hline
\end{tabular}
\caption{(Top) Vertex powers in the particle-hole channels. (Bottom) Vertex powers in the particle-particle channels.}
\label{T:A1}
\end{table}

\begin{figure}[h]
\centering
\includegraphics[width=8cm]{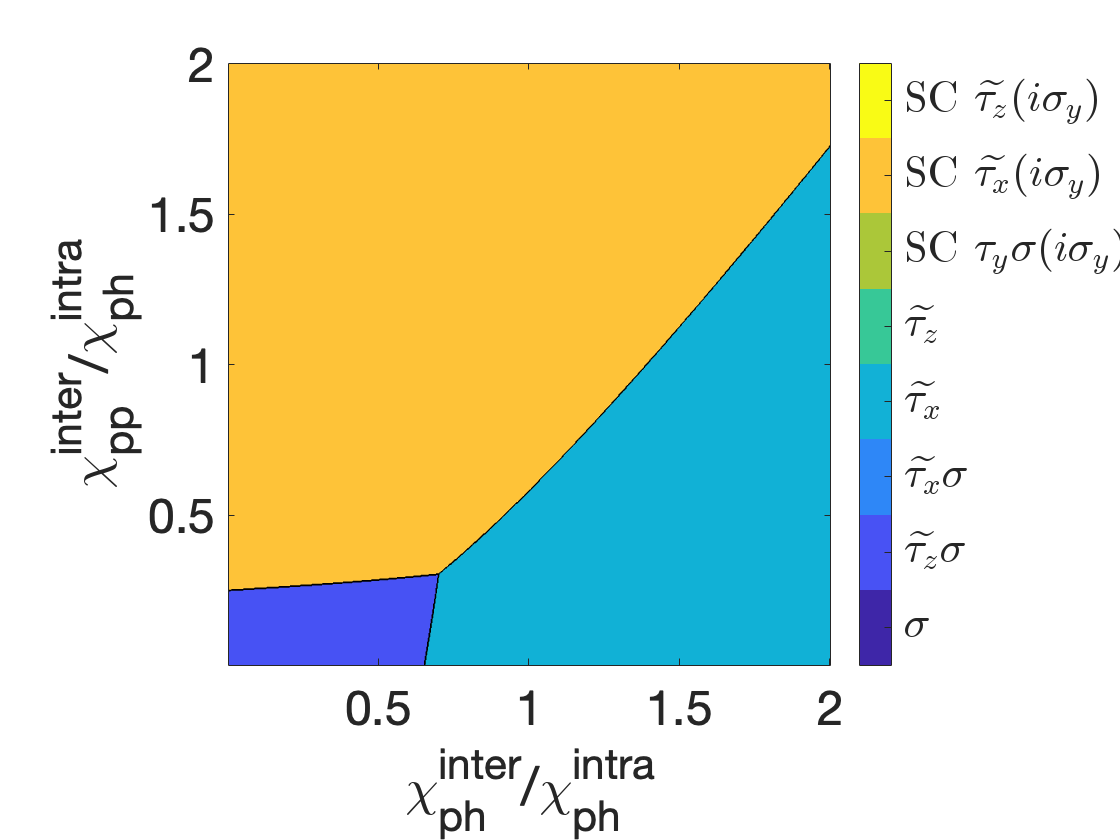}
\includegraphics[width=8cm]{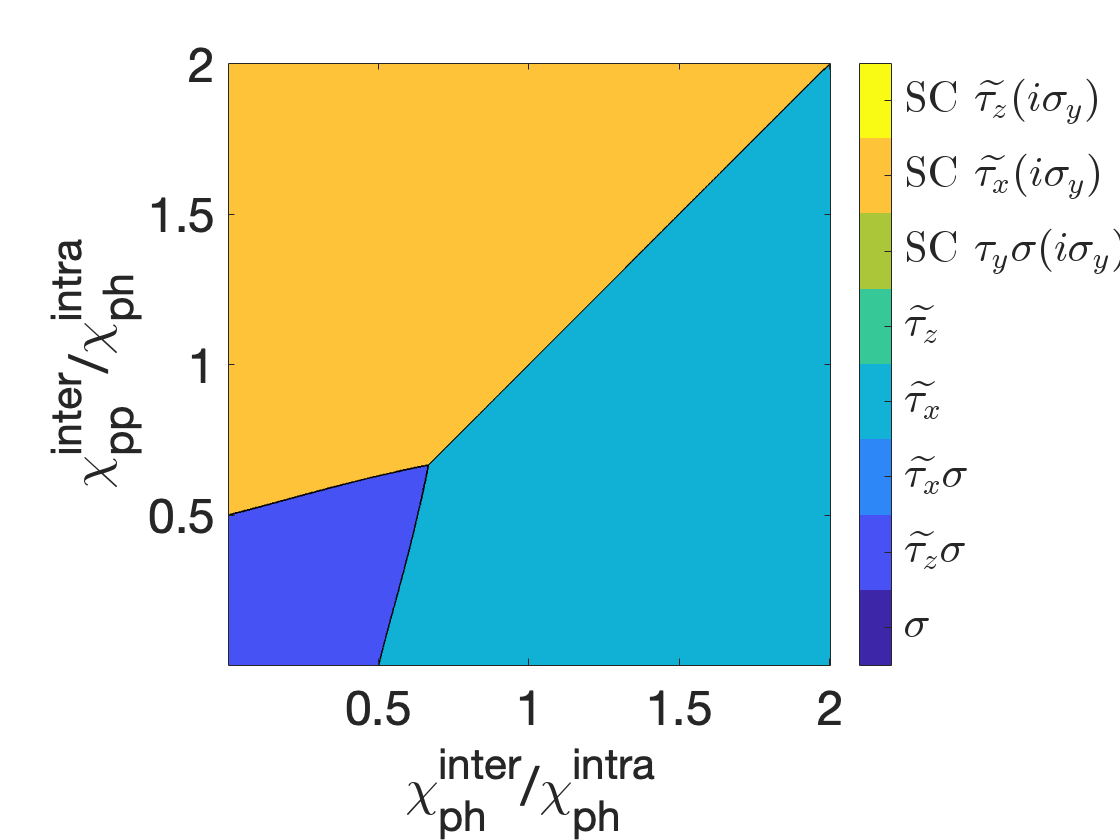}
\caption{(Left) Instabilities at the stable fixed point, as a function of the bare susceptibility ratio. (Right) The corresponding fixed point solution with $g_{1,2}<0$, when $g_5$ is not considered. }
\label{F:A1}
\end{figure}

The stable fixed point solution is shown in the left panel of Fig.\ref{F:A1}. We omit the trivial fixed point, where all interactions vanish. {\red We numerically solve the RG flow equations and show the leading instabilities for various bare susceptibility ratios.} The three possible instabilities are inter-band SC $\Delta=\widetilde{\tau_x}i\sigma_y$ (yellow), nematicity $O=\widetilde{\tau_x}$ (lighter blue), and altermagnetism $M=\widetilde{\tau_z}\sigma_z$ (darker blue). They become dominant when $\chi_{pp}^{inter}$, $\chi_{ph}^{inter}$, and $\chi_{ph}^{intra}$ are dominant, respectively. This is consistent with the results in the main text. {\red We checked the $g_{3,4}$ flow at the fixed point, where $g_{1,2,5}$ further push both $g_4\pm g_3$ to repulsion. Due to the BCS corrections within $g_4\pm g_3$, they should then remain small.}

\begin{figure}[ht]
\centering
\includegraphics[width=4cm]{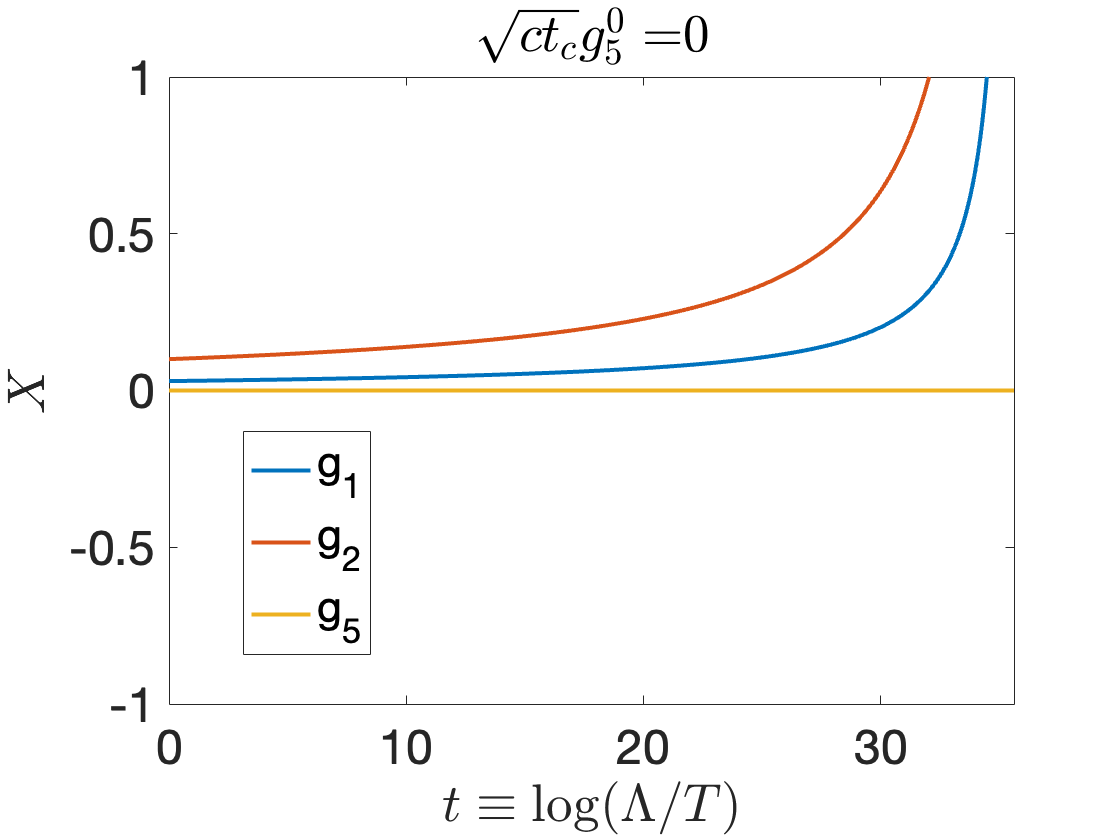}
\includegraphics[width=4cm]{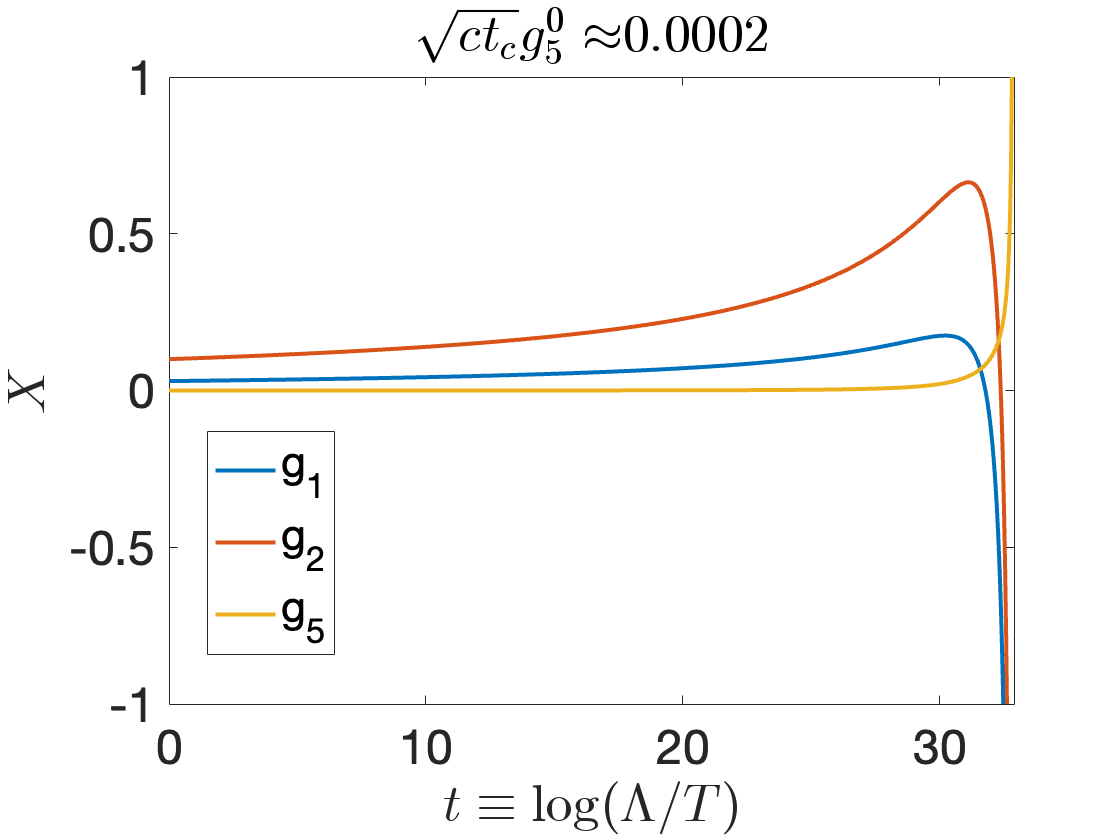}
\caption{RG flow for initial interactions $a=0.5$, $b=0.2$, $g_1=0.03$ and $g_2=0.1$. (left) For $g_5=0$, $g_{1,2}$ flows to $+\infty$. (right) A small $g_5\neq0$ can drive $g_{1,2}$ to $-\infty$. }
\label{F:A2}
\end{figure}

Qualitatively, $g_5$ does not introduce a new fixed point solution. The fixed point $g_{1,2}<0$ exists even if $g_5$ is not considered (Right panel of Fig.\ref{F:A1}). However, $g_5$ leads to negative BCS corrections to $g_{1,2}$, and makes this fixed point more preferable.
An example is shown in Fig. \ref{F:A2}. When $g_5$-interaction is absent, $g_1$ and $g_2$ flow to positive infinity for the given initial interactions. As $g_2$ approaches 1, the one-loop RG flow becomes uncontrolled and this leads to an instability. The above fixed point is unstable to a non-zero $g_5$-interaction. As shown in the right panel, {\red interestingly, even a tiny $\widetilde{g_5}=\sqrt{ct_c}g_5$ can drive $g_1$ and $g_2$ to negative infinity, before the theory becomes uncontrolled. This is because the RG flow of $g_5$ makes it grow exponentially before coming close to the fixed point.}

\section{Effect of SOC}
We now consider the orthorhombic $D_{2h,2}$ class with SOC, which is relevant for L44(pbam). It has the following kp Hamiltonian\cite{suh:2023} at the $(\pi,\pi)$ point:
\begin{equation}
\begin{split}
H=t_1k_xk_y\tau_x+t_2k_xk_y\tau_z+\lambda_zk_xk_y\tau_y\sigma_z,
\end{split}
\end{equation}
{\red Here, operators $\tau_x$, $\tau_z$, and $\tau_y\sigma_z$ all share the same symmetry as $k_xk_y$. The Hamiltonian can be diagonalized by the unitary transformation $u\propto c_1I+c_2\tau_x\sigma_z+c_3\tau_z\sigma_z+c_4\tau_y$, with k-independent prefactors $c_{1,2,3,4}$. Under such simple k-independent rotation, operators $\widetilde{\tau_x}$, $\widetilde{\tau_z}$, and $\widetilde{\tau_y\sigma_z}$ in the band basis should still share the same symmetry as $k_xk_y$.}

In this class, all three terms share the same symmetry as $k_xk_y$, so there are more allowed interactions. {\red We again use Cooper pairs to derive interactions. The three Cooper pairs below share the same symmetry:} $\widetilde{\tau_xi\sigma_y}:\Delta_{xy,1}=c_{2\downarrow}c_{1\uparrow}+c_{1\downarrow}c_{2\uparrow}$, $\widetilde{\tau_zi\sigma_y}:\Delta_{xy,2}=c_{1\downarrow}c_{1\uparrow}-c_{2\downarrow}c_{2\uparrow}$, and $\widetilde{\tau_y\sigma_zi\sigma_y}:\Delta_{xy,3}=i(c_{1\downarrow}c_{2\uparrow}+c_{2\downarrow}c_{1\uparrow})$. {\red Multiplying these Fermionic bilinears, we have three interactions}
\begin{equation}
\begin{split}
&g_5(c^\dagger_{2\downarrow}c^\dagger_{1\uparrow}+c^\dagger_{1\downarrow}c^\dagger_{2\uparrow})(c_{1\downarrow}c_{1\uparrow}-c_{2\downarrow}c_{2\uparrow})+c.c.\\
&ig_6(c^\dagger_{1\downarrow}c^\dagger_{2\uparrow}+c^\dagger_{2\downarrow}c^\dagger_{1\uparrow})(c_{1\downarrow}c_{1\uparrow}-c_{2\downarrow}c_{2\uparrow})+c.c.\\
&ig_7(c_{1\downarrow}^\dagger c_{2\uparrow}^\dagger+c_{2\downarrow}^\dagger c_{1\uparrow}^\dagger)(c_{2\downarrow}c_{1\uparrow}-c_{2\uparrow}c_{1\downarrow})+c.c..
\end{split}
\end{equation}
Notably, $g_7$ is different from the exchange interaction. The absence of spin-rotational symmetry also allows another inter-patch Hubbard interaction: $g_8(n_{1\uparrow}n_{2\uparrow}+n_{1\downarrow}n_{2\downarrow})$, and the spin-flip interactions:
$g_9c_{1\uparrow}^\dagger c_{2\uparrow}^\dagger c_{2\downarrow} c_{1\downarrow}+c.c.$.

In total, 9 interactions are needed under SOC for the orthorhombic class $D_{2h,2}$. They are exchange interaction $g_1$, inter-band Hubbard interaction $g_2n_{1\uparrow}n_{2\downarrow}$, pair-hopping interaction $g_3$, intra-band Hubbard interaction $g_4$, real single-hopping interaction $g_5$, imaginary single-hopping interaction $ig_6$, imaginary exchange interaction $ig_7$, inter-band Hubbard interaction $g_8n_{1\uparrow}n_{2\uparrow}$,  and spin-flip interaction $g_9$. 

The necessity of these nine interactions can also be understood from the point group classification at the high-symmetry point: $A_g + 2B_{1g} + [A_g]$. These group elements correspond to operators $1$, $(\tau_x,\tau_z)$, and $\tau_y$.
When including the spin to the antisymmetric (time-reversal odd) element $[A_g]$, the point groups elements are $A_g + 3B_{1g} + B_{2g}+B_{3g}$. This corresponds to operators $1$, $(\tau_x,\tau_z,\tau_y\sigma_z)$, $\tau_y\sigma_x$ and $\tau_y\sigma_y$.
Here, 3 elements have the same symmetry $B_{1g}=k_xk_y$. Their couplings generate 6 interactions. The other three independent elements generate the other 3 interactions.

In the main text, we neglect SOC and assume spin-rotational symmetry. Then operators $1$, $(\tau_x,\tau_z)$ generate 4 interactions, while operators $\tau_y\sigma_{1,2,3}$ generate interactions with the same strength. In total 5 interactions are needed.

\section{Anomalous Hall effect for orbital altermagnetic state}
{\red In this section, we will explain the anomalous Hall effect for the orbital altermagnetic state under $\epsilon_{xy}$ strain. The anomalous Hall effect requires the breaking of time-reversal symmetry and reflection symmetry $\widetilde{M_{x,y}}$. These symmetry breakings must be detectable in real space, not just in spin space. This is why SOC is necessary for ferromagnetism to exhibit anomalous Hall effect.

\begin{figure}[h]
\centering
\includegraphics[width=6cm]{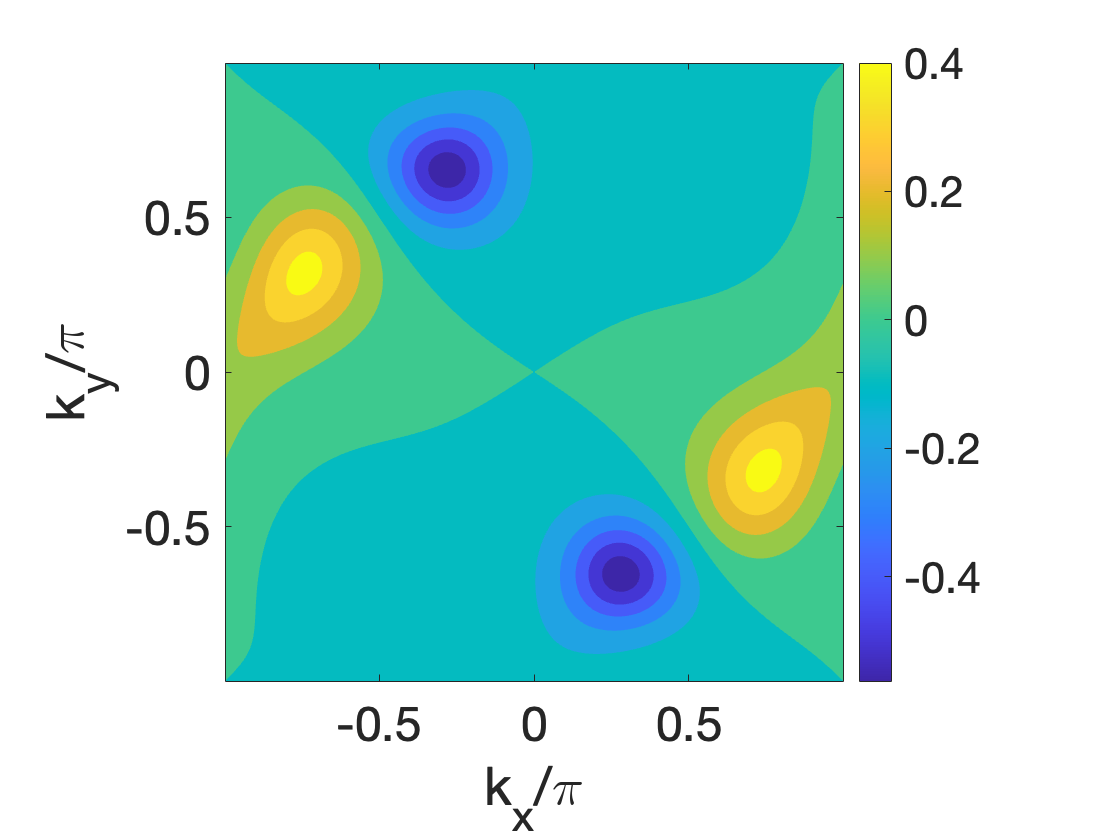}
\caption{Berry curvature for the lower-band of Eq.\ref{Eq:berry}, with $4t_4=t_5=M=\epsilon_{xy}=1$, and $r=0.5$. {\red On an orthorhombic Fermi surface, the averaged Berry curvature is generically non-zero.} }
\label{F:A3}
\end{figure}

The orbital altermagnetic state is a { real space} current-loop order that breaks time-reversal symmetry while maintaining all crystal symmetries. As the $\epsilon_{xy}$ strain breaks the { real space} mirror symmetry $\widetilde{M_{x,y}}$, the anomalous Hall conductivity will be non-zero. The anomalous Hall effect here is solely a real-space feature and does not require SOC. Since spin is not relevant to the following discussion, we will consider a spinless Hamiltonian:}
\begin{equation}
\begin{split}
H&=4t_4 \cos\frac{k_x}{2}\cos\frac{k_y}{2}(1+r\cos k_x)\tau_x\\
&+t_5 \sin k_x\sin k_y \tau_z+M\sin\frac{k_x}{2}\sin\frac{k_y}{2}\tau_y+\epsilon_{xy}\tau_z
\end{split}
\label{Eq:berry}
\end{equation}
The above Hamiltonian is centered at $\Gamma$ point. At the $(\pi,\pi)$ point, the orbital altermagnetic state reduces to $M\tau_y$, as written in the main text. The $\tau_z$ term captures the same symmetry as the $\epsilon_{xy}$-strain. {\red The Berry curvature is shown in Fig.\ref{F:A3}. For orthorhombic systems, we have included the symmetry-allowed $r$ term to explicitly break the 4-fold rotational symmetry. On any orthorhombic Fermi surface, the averaged Berry curvature is generally non-zero. This results in a non-zero anomalous Hall conductivity. }

{\red
\section{discussions on g-wave altermagnetism}
In this work, we have focused on coincident Van Hove singularities and the new interaction. At this specific Van Hove momentum, excitations can only be formed through band degrees of freedom, represented by the $\tau_{1,2,3}$ matrices. The only valid magnetic excitations are $\sigma_i$, $\tau_1\sigma_i$, and $\tau_3\sigma_i$. Since both $\tau_{1,3}$ possess $k_xk_y$ symmetry, they result in d-wave altermagnetism. g-wave altermagnetism is irrelevant at the Van Hove momentum point. For instance, g-wave magnetism $k_xk_y(k_x^2-k_y^2)\sigma_i$ can be expressed as $(k_x^2-k_y^2)\tau_1\sigma_i$ or $(k_x^2-k_y^2)\tau_3\sigma_i$, both of which vanish at the Van Hove momentum point. This indicates that the specific dispersion of the kp Hamiltonian does not support g-wave or i-wave altermagnetism at the Van Hove point.

However, more generally, if we study the more conventional Van Hove problem, it is possible to include those high-order altermagnetic states as excitations at the Van Hove point. For example, a relevant kp Hamiltonian\cite{suh:2023} is:
$H=t_0k^2+t_3k_xk_y\tau_3+t_1k_xk_y(k_x^2-k_y^2)\tau_1$. Here, $\tau_1$ exhibits g-wave symmetry $k_xk_y(k_x^2-k_y^2)$, while $\tau_3$ exhibits d-wave symmetry $k_xk_y$. This Hamiltonian describes a Van Hove singularity if the $\tau_3$ term dominates the $t_0$ term. Both bands host a Van Hove singularity at the band crossing, similar to this work. However, the new interaction $g_5$ will not appear because $\tau_1$ and $\tau_3$ have different symmetries. This Van Hove singularity can support g-wave altermagnetism, which does not vanish at the Van Hove point, in the form of $\tau_1\sigma_i$. For states near the band crossing, $\tau_3$ labels the two bands. The $\tau_1\sigma_i$ state is then a inter-patch magnetic states. This state can be mapped to the antiferromagnetism in cuprates, which is also a inter-patch state between the Van Hove singularities at $(0,\pi)$ and $(\pi,0)$. Such state can be stabilized if the two Van Hove singularities exhibit additional nesting.

\end{document}